\documentclass[10pt,journal,compsoc]{IEEEtran}
\makeatletter\if@twocolumn\PassOptionsToPackage{switch}{lineno}\else\fi\makeatother

\usepackage{graphicx}
\usepackage[T1]{fontenc}
\usepackage{multirow}

\usepackage{booktabs,caption}
\usepackage{threeparttable}

\ifCLASSOPTIONcompsoc
  \usepackage[nocompress]{cite}
\else
  \usepackage{cite}
\fi
\ifCLASSINFOpdf
\else
\fi
\usepackage{amsmath}
\usepackage[cmintegrals]{newtxmath}
\usepackage{url,multirow,morefloats,floatflt,cancel,tfrupee}
\makeatletter

\AtBeginDocument{\@ifpackageloaded{textcomp}{}{\usepackage{textcomp}}}
\makeatother
\usepackage{colortbl}
\usepackage{xcolor}
\usepackage{pifont}
\usepackage[nointegrals]{wasysym}
\makeatletter

\def\mcWidth#1{\csname TY@F#1\endcsname+\tabcolsep}

\def\cAlignHack{\rightskip\@flushglue\leftskip\@flushglue\parindent\z@\parfillskip\z@skip}
\def\rAlignHack{\rightskip\z@skip\leftskip\@flushglue \parindent\z@\parfillskip\z@skip}

\@ifundefined{etal}{}{}

\usepackage{ifxetex}
\ifxetex\else\if@twocolumn\@ifpackageloaded{stfloats}{}{\usepackage{dblfloatfix}}\fi\fi

\AtBeginDocument{
\expandafter\ifx\csname eqalign\endcsname\relax
\def\eqalign#1{\null\vcenter{\def\\{\cr}\openup\jot\m@th
  \ialign{\strut$\displaystyle{##}$\hfil&$\displaystyle{{}##}$\hfil
      \crcr#1\crcr}}\,}
\fi
}

\AtBeginDocument{%
  \@ifpackageloaded{endfloat}%
   {\renewcommand\efloat@iwrite[1]{\immediate\expandafter\protected@write\csname efloat@post#1\endcsname{}}}{\newif\ifefloat@tables}%
}%

\def\BreakURLText#1{\@tfor\brk@tempa:=#1\do{\brk@tempa\hskip0pt}}
\let\lt=<
\let\gt=>
\def\processVert{\ifmmode|\else\textbar\fi}

\@ifundefined{subparagraph}{
\def\subparagraph{\@startsection{paragraph}{5}{2\parindent}{0ex plus 0.1ex minus 0.1ex}%
{0ex}{\normalfont\small\itshape}}%
}{}

\newcommand\role[1]{\unskip}
\newcommand\aucollab[1]{\unskip}
  
\@ifundefined{tsGraphicsScaleX}{\gdef\tsGraphicsScaleX{1}}{}
\@ifundefined{tsGraphicsScaleY}{\gdef\tsGraphicsScaleY{.9}}{}
\def\checkGraphicsWidth{\ifdim\Gin@nat@width>\linewidth
	\tsGraphicsScaleX\linewidth\else\Gin@nat@width\fi}

\def\checkGraphicsHeight{\ifdim\Gin@nat@height>.9\textheight
	\tsGraphicsScaleY\textheight\else\Gin@nat@height\fi}

\def\fixFloatSize#1{}
\let\ts@includegraphics\includegraphics

\def\inlinegraphic[#1]#2{{\edef\@tempa{#1}\edef\baseline@shift{\ifx\@tempa\@empty0\else#1\fi}\edef\tempZ{\the\numexpr(\numexpr(\baseline@shift*\f@size/100))}\protect\raisebox{\tempZ pt}{\ts@includegraphics{#2}}}}

\AtBeginDocument{\def\includegraphics{\@ifnextchar[{\ts@includegraphics}{\ts@includegraphics[width=\checkGraphicsWidth,height=\checkGraphicsHeight,keepaspectratio]}}}

\DeclareMathAlphabet{\mathpzc}{OT1}{pzc}{m}{it}

\def\URL#1#2{\@ifundefined{href}{#2}{\href{#1}{#2}}}

\def\UrlOrds{\do\*\do\-\do\~\do\'\do\"\do\-}%
\g@addto@macro{\UrlBreaks}{\UrlOrds}

\edef\fntEncoding{\f@encoding}

\makeatother

\newif\ifmultipleabstract\multipleabstractfalse%
\makeatletter
\AtBeginDocument{\@ifpackageloaded{longtable}{%
\def\LT@makecaption#1#2#3{%
  \LT@mcol\LT@cols c{\hbox to\z@{\hss\parbox[t]\LTcapwidth{%
    \sbox\@tempboxa{#1{#2: } #3}%
    \ifdim\wd\@tempboxa>\hsize
      #1{#2: }\textsc{#3}%
    \else
      \hbox to\hsize{\hfil\box\@tempboxa\hfil}%
    \fi
    \endgraf\vskip\baselineskip}%
  \hss}}}
}{}}
\makeatother

\usepackage{scalerel}
\usepackage{tikz}
\usetikzlibrary{svg.path}

\definecolor{orcidlogocol}{HTML}{A6CE39}
\tikzset{
  orcidlogo/.pic={
    \fill[orcidlogocol] svg{M256,128c0,70.7-57.3,128-128,128C57.3,256,0,198.7,0,128C0,57.3,57.3,0,128,0C198.7,0,256,57.3,256,128z};
    \fill[white] svg{M86.3,186.2H70.9V79.1h15.4v48.4V186.2z}
                 svg{M108.9,79.1h41.6c39.6,0,57,28.3,57,53.6c0,27.5-21.5,53.6-56.8,53.6h-41.8V79.1z M124.3,172.4h24.5c34.9,0,42.9-26.5,42.9-39.7c0-21.5-13.7-39.7-43.7-39.7h-23.7V172.4z}
                 svg{M88.7,56.8c0,5.5-4.5,10.1-10.1,10.1c-5.6,0-10.1-4.6-10.1-10.1c0-5.6,4.5-10.1,10.1-10.1C84.2,46.7,88.7,51.3,88.7,56.8z};
  }
}

\newcommand\orcidicon[1]{\href{https://orcid.org/#1}{\mbox{\scalerel*{
\begin{tikzpicture}[yscale=-1,transform shape]
\pic{orcidlogo};
\end{tikzpicture}
}{|}}}}

\usepackage{hyperref}
\usepackage{stmaryrd}
\usepackage{algpseudocode}
\usepackage{cleveref}
\usepackage{algorithm}

\newcommand{\algalign}[2]

\let\oldemptyset\emptyset
\let\emptyset\varnothing

\usepackage{pifont}
\newcommand{\xmark}{\ding{55}}

\newtheorem{definition}{Definition}
\newtheorem{example}{Example}

\begin{document}

%
\title{An Automatic Attribute Based Access Control Policy Extraction from Access Logs}

%
%
\author{Leila~Karimi$^{\textsuperscript{\orcidicon{0000-0002-4346-728X}}}$,~\IEEEmembership{Student~Member,~IEEE,}
        Maryam~Aldairi$^{\textsuperscript{\orcidicon{0000-0002-5265-1281}}}$,~\IEEEmembership{Student~Member,~IEEE,}
        James~Joshi$^{\textsuperscript{\orcidicon{0000-0003-4519-9802}}}$,~\IEEEmembership{Senior~Member,~IEEE,}
        and~Mai~Abdelhakim$^{\textsuperscript{\orcidicon{0000-0001-8442-0974}}}$,~\IEEEmembership{Member,~IEEE}
\thanks{L. Karimi, M. Aldairi, and J. Joshi are with the School of Computing and Information, University of Pittsburgh.}
\thanks{M. Abdelhakim is with Electrical and Computer Engineering, Swanson School of Engineering, University of Pittsburgh.}
\thanks{Email addresses: \{leila.karimi, ma.aldairi, jjoshi, and maia\}@pitt.edu}

\thanks{© 2021 IEEE. Personal use of this material is permitted. Permission from IEEE must be obtained for all other uses, including reprinting/republishing this material for advertising or promotional purposes, collecting new collected works for resale or redistribution to servers or lists, or reuse of any copyrighted component of this work in other works.}
}

\IEEEtitleabstractindextext{
\begin{abstract}
With the rapid advances in computing and information technologies, traditional access control models have become inadequate in terms of capturing fine-grained, and expressive security requirements of newly emerging applications. An attribute-based access control (ABAC) model provides a more flexible approach to addressing the authorization needs of complex and dynamic systems. While organizations are interested in employing newer authorization models, migrating to such models pose as a significant challenge. Many large-scale businesses need to grant authorizations to their user populations that are potentially distributed across disparate and heterogeneous computing environments. Each of these computing environments may have its own access control model. The manual development of a single policy framework for an entire organization is tedious, costly, and error-prone. 

In this paper, we present a methodology for automatically learning ABAC policy rules from access logs of a system to simplify the policy development process. The proposed approach employs an unsupervised learning-based algorithm for detecting patterns in access logs and extracting ABAC authorization rules from these patterns. In addition, we present two policy improvement algorithms, including rule pruning and policy refinement algorithms to generate a higher quality mined policy. Finally, we implement a prototype of the proposed approach to demonstrate its feasibility. 
\end{abstract}
}

\maketitle 
      
\IEEEdisplaynontitleabstractindextext

%
\IEEEpeerreviewmaketitle

\section{Introduction}
\IEEEPARstart{A}{ccess} control systems are critical components of information systems that help protect information resources from unauthorized accesses. Various access control models and approaches have been proposed in the literature including Discretionary Access Control (DAC) \cite{sandhu1994access} \cite{harrison1976protection}, Mandatory Access Control (MAC) \cite{bell1973secure} \cite{sandhu1993lattice}, and Role-Based Access Control (RBAC) \cite{sandhu1996role}. However, with the rapid advances in newer computing and information technologies (e.g., social networks, Internet of Things (IoT), cloud/edge computing, etc.), existing  access control (AC) approaches have become inadequate in providing flexible and expressive authorization services \cite{fong2011relationship}. For example, a health care environment requires a more expressive AC model that meets the needs of patients, health care providers as well as other stakeholders in the health care ecosystem \cite{jin2009patient, karimi2017multi}.  \emph{Attribute Based Access Control} (ABAC) models present a promising approach that addresses newer challenges in emerging applications \cite{hu2013guide}. An ABAC approach grants access rights to users based on attributes of entities in the system (i.e., user attributes, object attributes, and environmental conditions) and a set of authorization rules. 

Although organizations and developers are interested in employing the next generation AC models, adopting such policy frameworks poses a significant challenge. Many large organizations need to grant authorization to their vast user populations distributed across disparate computing environments, including legacy systems. Each of these computing environments may have its own AC model. The manual development of a single policy for the entire organization is tedious and error-prone. \emph{Policy Mining} techniques have been proposed in the literature to address such challenges to help organizations cut the cost, time, and error of policy development/management. Policy mining algorithms ease the migration to more recent/appropriate authorization models by completely (or partially) automating the process of constructing AC policies. 

Policy mining techniques were first introduced for developing RBAC policies. Kuhlmann \textit{et al.} coined the term ``role mining" to refer to a data mining approach that constructs roles from a given permission assignment dataset \cite{kuhlmann2003role}; this work was followed by various role mining techniques, such as \cite{schlegelmilch2005role, molloy2008mining, xu2012algorithms}. Although the proposed approaches are beneficial in developing optimal sets of roles, they are not applicable in extracting ABAC policies. 

Xu and Stoller were the first to study the problem of mining ABAC policies from given access control matrices
or logs \cite{xu2014mining, xu2015mining}. Following that, several researchers have investigated various ABAC policy mining techniques \cite{medvet2015evolutionary, iyer2018mining, cotrini2018mining}. However, these studies suffer from several limitations, as follows:

\begin{itemize}
    \item First, the existing approaches do not support mining authorization rules with negative filters. An ABAC policy rule can be comprised of a set of positive and negative filters. Negative filters are useful in scenarios when an exception needs to be expressed. For example, a healthcare provider can express the following rule using a negative attribute filter:
    ``\textit{A nurse can read a patient's record except for payment purposes}."
    Using negative filters in rule expressions results in a more concise authorization policy (Section \ref{evaluation}). 
    
    \item Second, some proposed approaches such as in \cite{xu2014mining, xu2015mining, iyer2018mining} are unable to mine a high-quality policy when the given access log is not complete in the sense that every possible combination of attribute values is not included in the access log (Section \ref{problem}). 
    
    \item Third, the proposed approaches are unable to mine a policy from noisy access logs containing over-assignments and under-assignments \cite{medvet2015evolutionary, cotrini2018mining}. Having noisy access records is a common problem in evolving domains such as IoT or social networks \cite{marinescu2017ivd}. It is essential that an ABAC policy miner should be capable of handling a reasonable amount of noise to be applicable to real-world applications.

    \item Last but not the least, the existing approaches do not include techniques for improving the mined policy after the first round of policy extraction. In addition, in scenarios where the authorization policies may change over time (such as in social networks with addition and removal of various applications), these approaches do not provide any guidelines for adjusting the policy. This makes practical deployment of these approaches very difficult.
\end{itemize}

Furthermore, none of the existing work addresses these issues in an integrated way. In this paper, we propose a machine learning based ABAC policy mining approach to address these challenges.
To summarize, the primary contributions of this paper are as follows:
\begin{enumerate}
    \item We propose an unsupervised learning based approach to extract ABAC policy rules that contain both positive and negative attribute filters as well as positive and negative relation conditions.
    \item The proposed policy mining approach is effective even with an incomplete set of access logs and in presence of noise. 
    \item As part of the unsupervised learning based approach, we propose the rule pruning and policy refinement algorithms to enhance the quality of the mined policy and to ease its maintenance. 
    \item We propose a \emph{policy quality metric} based on policy correctness and conciseness to be able to compare different sets of mined policy rules and to select the best one based on some given criteria. 
    \item We implement a prototype of the proposed model and evaluate it using various ABAC policies to show its efficiency and effectiveness. 
\end{enumerate}

To the best of our knowledge, our proposed approach is the first unsupervised learning based ABAC policy mining method that can be used to extract ABAC policies with both positive and negative attribute and relationship filters.



The rest of the paper is organized as follows. In Section \ref{preliminaries}, we overview the ABAC model and its policy language as well as the unsupervised learning algorithm. In Section \ref{problem}, we define the ABAC policy extraction problem, discuss the related challenges, and introduce the metrics for evaluating the extracted policy. In Section \ref{proposed}, we present the proposed ABAC policy extraction approach. In Section \ref{evaluation}, we present the evaluation of the proposed approach on various sets of policies. We present the related work in Section \ref{relatedwork} and the conclusions and future work in Section \ref{conclusion}. 
    
\section{Preliminaries}\label{preliminaries}
In this section, we overview ABAC, the ABAC policy language, and the unsupervised learning algorithm.

\subsection{ABAC Model}
In 2013, NIST published a ``\textit{Guide to ABAC Definition and Consideration}" \cite{hu2013guide},  according to which, ``\textit{the ABAC engine can make an access control decision based on the assigned attributes of the requester, the assigned attributes of the object, environment conditions, and a set of policies that are specified in terms of those attributes and conditions}.'' Throughout the paper, we use \emph{user attributes}, \emph{object attributes}, and \emph{session attributes} to refer to the attributes of the requester, attributes of the object, and the environmental attributes/conditions, respectively. 

Accordingly, $U$, $O$, $S$, $OP$ are sets of users, objects, sessions, and operations in a system and user attributes ($A_u$), object attributes ($A_o$), and session attributes ($A_s$) are mappings of subject attributes, object attributes, and environmental attributes as defined in the NIST Guide \cite{hu2013guide}. $E = U \cup O \cup S$ and $A = A_u \cup A_o \cup A_s$ are the sets of all entities and all attributes in the system, respectively.

\begin{definition}
(\textbf{Attribute Range}). 
Given an attribute $a \in A$, the \emph{attribute range} $V_a$ is the set of all valid values for $a$ in the system.
\end{definition}

\begin{definition}
(\textbf{Attribute Function}).
Given an entity $e \in E$, an \emph{attribute function} $f_{a\_e}$ is a function that maps an entity to a specific value from the attribute range. Specifically, $f_{a\_e}(e, a)$ returns the value of attribute $a$ for entity $e$. 
\end{definition}

\begin{example}\label{ex1}
$f_{a\_e}(John, position) = \mathit{faculty}$ indicates that the value of attribute \emph{position} for user \emph{John} is \emph{faculty}.
\end{example}

\begin{example}\label{ex2}
$f_{a\_e}(dep1, crs) = \{cs101, cs601, cs602\}$ indicates that the value of attribute \emph{crs} for object \emph{dep1} is a set $\{cs101, cs601, cs602\}$.
\end{example}

Each attribute in the system can be a single-valued (atomic) or multi-valued (set). In Example \ref{ex1} \emph{position} is a single-valued attribute while \emph{crs} is a multi-valued attribute in Example \ref{ex2}. For simplicity, we consider only atomic attributes in this work. Actually, the process of extracting ABAC policy with multi-valued attributes is exactly the same as that with atomic attributes, however, we need to pre-process data to convert each multi-valued attribute to a set of atomic attributes. This can be done using various techniques such as defining dummy variables \cite{suits1957use}, 1-of-$K$ scheme \cite{bishop2006pattern}, etc. At the end of the process and when policy rules are extracted, we need one more step to convert back atomic attribute filters to the corresponding multi-valued attribute filters.

Attribute filters are used to denote the sets of users, objects, and sessions to which an authorization rule applies.

\begin{definition}
(\textbf{Attribute Filter}). 
An \emph{attribute filter} is defined as a set of tuples $\mathcal{F} = \{ \langle a, v | !v \rangle | \: a \in A$ and $v \in V_a \}$. Here $\langle a, v \rangle $ is a positive attribute filter tuple that indicates $a$  has value $v$, and $ \langle a, !v \rangle $ is a negative attribute filter tuple that indicates $a$ has any value in its range except $v$.
\end{definition}

\begin{example}\label{ex3}
Tuple $\langle label, !top\text{-}secret \rangle $ points to all entities in the system that do not have ``\emph{top-secret}" as their security label ``\emph{label}".
\end{example}

\begin{definition}
(\textbf{Attribute Filter Satisfaction}). 
An entity $e \in E$ satisfies an attribute filter $\mathcal{F}$, denoted as $e \models \mathcal{F}$, iff 

\begin{equation*}
  \begin{gathered}
 \forall \langle {a}_i, {v}_i\rangle \: \in \mathcal{F} : f_{a\_e}(e, a_i) = {v}_i \: \land \\
\forall \langle {a}_i, !{v}_i\rangle \: \in \mathcal{F} : f_{a\_e}(e, a_i) \neq {v}_i.
  \end{gathered}
\end{equation*}
\end{definition}

\begin{example}
Suppose $A_u = \{dept, position, courses\}$. The set of tuples $\mathcal{F_U} = \{\langle dept, CS\rangle, \langle position, \allowbreak grad\rangle\}$ denotes a user attribute filter. Here, the graduate students in the CS department satisfy $\mathcal{F_U}$.
\end{example}

\begin{definition}
(\textbf{Relation Condition}). 
A \emph{relation condition} is defined as a set of tuples $\mathcal{R} = \{\langle a, b | !b\rangle | \: a, b \in A \land \: a \neq b \}$. Here $\langle a, b \rangle $ is a positive relation condition tuple that indicates $a$ and $b$ have the same values, and $ \langle a, !b \rangle $ is a negative relation condition tuple that indicates $a$ and $b$ do not have the same values.
\end{definition}

A relation is used in a rule to denote the equality condition between two attributes of users, objects, or sessions. Note that the two attributes in the relation condition must have the same range. 

\begin{definition}
(\textbf{Relation Condition Satisfaction}). 
An entity $e \in E$ satisfies a relation condition $\mathcal{R}$, denoted as $e \models \mathcal{R}$, iff 
\begin{equation*}
  \begin{gathered}
     \forall \langle{a}_i, {b}_i\rangle \: \in \mathcal{R} : f_{a\_e}(e, a_i) = f_{a\_e}(e, b_i) \: \\
     \forall \langle{a}_i, !{b}_i\rangle \: \in \mathcal{R} : f_{a\_e}(e, a_i) \neq f_{a\_e}(e, b_i).
    \end{gathered}
\end{equation*}
\end{definition}

\begin{definition}
(\textbf{Access Request}).
An \emph{access request} is a tuple $q = \langle u, o, s, op\rangle$ where user $u \in U$ sends a request to the system to perform operation $op \in OP$ on object $o \in O$ in session $s \in S$.
\end{definition}

\begin{definition}
(\textbf{Authorization Tuple/Access Log}).
An \emph{authorization tuple} is a tuple $t = \langle q, d\rangle$ containing decision $d$ made by the access control system for request $q$. An \emph{Access Log} $\mathcal{L}$ is a set of such tuples.
\end{definition}

The decision $d$ of an authorization tuple can be \emph{permit} or \emph{deny}. The tuple with \emph{permit} decision means that user $u$ can perform an operation $op$ on an object $o$ in session $s$. The authorization tuple with \emph{deny} decision means that user $u$ cannot perform operation $op$ on object $o$ in session $s$.

Access log is a union of \emph{Positive Access Log}, $\mathcal{L^+}$, and \emph{Negative Access Log}, $\mathcal{L^-}$, where:

$$\mathcal{L^+} = \{\langle q, d\rangle | \langle q, d\rangle \: \in \mathcal{L} \land d = permit \},$$
and
$$\mathcal{L^-} = \{\langle q, d\rangle  | \langle q, d\rangle \: \in \mathcal{L} \land d = deny \}.$$

\begin{definition}
(\textbf{ABAC Rule}).
An \emph{access rule} $\rho$ is a tuple $ \langle \mathcal{F}, \mathcal{R}, op | !op\rangle $, where $\mathcal{F}$ is an attribute filter, $\mathcal{R}$ is a relation condition, and $op$ is an operation. $!op$ is a negated operation that indicates the operation can have any value except $op$.
\end{definition}

\begin{example}
Consider rule ${\rho}_1 = \langle \{\langle position,student\rangle, \: \: \allowbreak \langle location, campus\rangle, \langle type, article\rangle\}, \{\langle dept_u, dept_o \rangle \},\: \: \allowbreak read\rangle$. It can be interpreted as ``\textit{A student can read an article if he/she is on campus and his/her department matches the department of the article}".
\end{example}

\begin{definition}
(\textbf{Rule Satisfaction})
An access request $q = \langle u, o, s, op \rangle $ is said to satisfy a rule $\rho$, denoted as $q \models \rho$, iff 
\begin{equation*}
    \langle u, o, s \rangle \models \mathcal{F} \land \langle u, o, s \rangle \models \mathcal{R} \land op_{q} = op_{\rho}.
\end{equation*}
\end{definition}




\begin{definition}
(\textbf{ABAC Policy}).
An ABAC policy is a tuple $\pi = \langle E, OP, A, f_{a\_e}, \mathcal{P} \rangle $ where $E$, $OP$, $A$, and $\mathcal{P}$ are sets of entities, operations, attributes, and ABAC rules in the system and $f_{a\_e}$ is the attribute function.
\end{definition}

\begin{definition}
(\textbf{ABAC Policy Decision}).
The decision of an ABAC policy $\pi$ for an access request $q$ denoted as $d_\pi(q)$ is \emph{permit} iff:
$$\exists \rho \in \pi :  q \models \rho $$
otherwise, the decision is \emph{deny}.
\end{definition}

If an access request satisfies a rule of the access control policy, then the decision of the system for such access request is \textit{permit}. If the access request does not satisfy any rule in the access control policy then the decision of the system for such access request is \textit{deny}. 


TABLE \ref{tab:notations} summarizes the notations used in this paper.

\begin{table*}
\centering
\caption{Notations} \label{tab:notations}
\begin{tabular}{cl}  
\toprule
    Notation & Definition \\
\midrule
$U$,$O$, $S$, $OP$ & Sets of users, objects, sessions, and operations \\
$A_u$, $A_o$, and $A_s$ & Sets of user attributes, object attributes, and session attributes \\
$E = U \cup O \cup S$ & Set of all entities \\
$A = A_u \cup A_o \cup A_s$ & Set of all attributes \\
$V_a$ & Attribute Range: set of all valid values for $a \in A$ \\
$f_{a\_e}(e, a)$ & Attribute Function: a function that maps an entity $e \in E$ to a value from $V_a$ \\
$\mathcal{F} = \{ \langle a, v | !v \rangle | \: a \in A \land v \in V_a \}$ & Attribute Filter \\
$\mathcal{R} = \{\langle a, b\rangle | \: a, b \in A \land \: a \neq b \land V_a = V_b\}$ & Relation Condition \\
$q = \langle u, o, s, op\rangle$ & Access Request \\
$t = \langle q, d\rangle$ & Authorization Tuple, showing decision $d$ made by the system for request $q$ \\
$\mathcal{L}$ & Access Log, set of authorization tuples \\ 
$\mathcal{L^+} = \{\langle q, d\rangle | \langle q, d\rangle \: \in \mathcal{L} \land d = permit \}$ & Positive Access Log \\ 
$\mathcal{L^-} = \{\langle q, d\rangle  | \langle q, d\rangle \: \in \mathcal{L} \land d = deny \}$ & Negative Access Log \\
$\rho = \langle \mathcal{F}, \mathcal{R}, op|!op\rangle$ & ABAC Rule \\
$\mathcal{P}$ & Set of all policy rules \\ 
$\pi = \langle E, OP, A, f_{a\_e}, \mathcal{P} \rangle$ & ABAC Policy \\ 
$d_\pi(q)$ & The decision of an ABAC policy $\pi$ for an access request $q$ \\
$TP_{\pi|\mathcal{L}}$, $FP_{\pi|\mathcal{L}}$, $TN_{\pi|\mathcal{L}}$, and $FN_{\pi|\mathcal{L}}$ & Relative  True  Positive, False Positive, True Negative, and False Negative Rates \\ 
$ACC_{\pi|\mathcal{L}}$ & Relative Accuracy Rate \\ 
$F{\text -}score_{\pi|\mathcal{L}}$ & Relative F-score \\ 
$WSC(\pi)$ & Weighted Structural Complexity of policy $\pi$ \\ 
$\mathcal{Q}_{\pi}$ & Policy Quality  Metric\\ 
\bottomrule
\end{tabular}
\end{table*}

\subsection{Unsupervised Learning Algorithm}
Unsupervised learning algorithms try to infer a function that describes the structure of unlabeled data. They are useful when no or very few labeled data is available. We leverage such methods for extracting ABAC policies from access logs. 

In particular, given a set of authorization tuples, we employ an unsupervised learning approach to mine and extract an \emph{ABAC policy} that has high quality. An unsupervised learning approach is suitable because there is no labeled data available for desired ABAC rules. ABAC policy extraction, in this case, can be considered as a mapping between authorization tuples to a set of clusters that are representative of the desired ABAC  rules. Such a mapping can be expressed as a function, $h : \mathcal{X} \to \mathcal{Y}$, where:

\begin{enumerate}
  \item $\mathcal{X}$ is a set of authorization tuples (i.e., access log).
  \item $\mathcal{Y}$ is a set of numbered labels (i.e., cluster labels, each cluster corresponding to a rule of the ABAC policy $\pi$).
 \end{enumerate}
 
The goal is then to learn the function $h$ with low clustering error and mine the desired policy that is high quality.

\section{Problem Definition} \label{problem}


\subsection{ABAC Policy Extraction Problem}
Although organizations are interested in employing an ABAC model, adopting it is a big challenge for them. The manual development of such a policy is tedious and error-prone. \emph{Policy Mining} techniques have been proposed to address such challenges in order to reduce the cost, time, and error of policy development/maintenance. ABAC policy mining algorithms ease the migration to the ABAC framework by completely (or partially) automating the development of ABAC policy rules. 

The primary input to a policy mining algorithm is the log of authorization decisions in the system. The log indicates authorization decision (i.e., permit or deny) for any given access request by a user of the system. For ABAC policy mining, such a log is accompanied by attributes of entities involved in the log entries. The goal of a policy mining algorithm is to extract ABAC policy rules from access logs that have high quality with respect to some quality metrics (e.g., policy size and correctness).

We define the ABAC policy extraction problem formally as follows: 

\begin{definition}
(\textbf{ABAC Policy Extraction Problem}).
Let $I = <E, OP, A, f_{a\_e}, \mathcal{L}>$, where the components are as defined earlier, then the \emph{ABAC policy extraction problem} is to find a set of rules $\mathcal{R}$ such that the ABAC policy $\pi = <E, OP, A, f_{a\_e}, \mathcal{R}>$ has high quality with respect to $\mathcal{L}$.
\end{definition}

\subsection{Challenges and Requirements}
For an ABAC policy extraction approach to be applicable to a wide range of real-world scenarios, we identify the following challenges and requirements: 

\begin{enumerate}
\item \textit{Correctness of Mined Policy}: The mined policy must be consistent with original authorization log in that the access decision of the mined policy must result in the same access decision of the log entry. An inconsistent extracted policy may result in situations in which an originally authorized access is denied (\textit{more restrictive}) or originally unauthorized access is permitted (\textit{less restrictive}) by the system.

\item \textit{Complexity of Mined Policy}: The policy mining algorithm should endeavor to extracting a policy that is as concise as possible. Since the policy rules need to be manipulated by human administrators, the more concise they are, the more manageable and easier to interpret they would be. In addition, succinct rules are desirable as they are easier to audit and manage.

\item \textit{Negative Attribute Filters}: The ABAC policy mining solution should support both positive and negative attribute filters which will result in more concise and manageable mined policy.

\item \textit{Relation Conditions}: The solution should support the extraction of relation conditions for policy mining in order to generate more concise and manageable mined policy.

\item \textit{Sparse Logs}: In real-world, the access log that is input to the policy mining algorithm may be sparse, representing only a small fraction of all possible access requests. The policy mining algorithm must be able to extract useful rules even from a sparse log.


\item \textit{Mining Negative Authorization Rules}: An ABAC policy can contain both positive and negative rules which permit or deny access requests, respectively. The use of negative rules is helpful in situations where specifying exceptions to more general rules is important. Including negative policy rules would help in generating a more concise ABAC policy. Thus, the policy mining algorithm should be able to extract both positive and negative authorization rules.

\item \textit{Noisy Authorization Log}: In the real world and with complex and dynamic information systems, it is possible to have a noisy authorization log consisting of over-assignments and under-assignments. These issues occur either due to a wrong configuration of the original authorization system or improper policy updates by administrators. The policy mining algorithm should be capable of extracting meaningful rules even in presence of an acceptable amount of noise in the input access log. 

\item \textit{Dynamic and Evolving Policies}: Modern information systems are often dynamic. The authorization needs of these systems and the attributes of the entities in the environment evolve rapidly. These changes will result in over-assignments or under-assignments. The proposed method should employ a mechanism to support the dynamicity of the information systems and their authorization policies and ease the maintenance of evolving systems.


\end{enumerate}

Our proposed approach addresses all the requirements except the sixth one. Table \ref{tab:existing_techniques} shows the challenges that are addressed by our proposed approach and how it improves upon the state-of-the-art policy mining techniques. In Section \ref{relatedwork}, we discuss the existing solutions in details.

\begin{table*}
\centering
\caption{State-of-the-art ABAC Rule Mining Techniques} \label{tab:existing_techniques}
\begin{tabular}{lccccc}
\toprule
&  Xu \textit{et al.} \cite{xu2015mining} & Medvet \textit{et al.} \cite{medvet2015evolutionary} & Iyer \textit{et. al} \cite{iyer2018mining} & Cotrini \textit{et al.} \cite{cotrini2018mining} & Our Proposed Approach \\  \midrule
Policy  Correctness & \checkmark &\checkmark & \checkmark & \checkmark & \checkmark \\  
Policy  Complexity & \checkmark & \checkmark & \checkmark & \checkmark & \checkmark \\  
Negative  Attribute  Filters & \xmark & \xmark & \xmark & \xmark & \checkmark \\  
Relation  Conditions & \checkmark & \checkmark & \checkmark & \xmark & \checkmark \\  
Sparse  Logs & \xmark & \checkmark & \xmark & \checkmark & \checkmark \\ 
Negative Authorization Rules & \xmark & \xmark & \checkmark & \xmark & \xmark  \\ 
Noisy  Authorization Log & \checkmark & \xmark & \xmark & \xmark & \checkmark \\ 
System  Dynamicity & \xmark & \xmark & \xmark & \xmark & \checkmark \\
\bottomrule
\end{tabular}
\end{table*}


\subsection{Evaluation Metrics}
One of the main metrics for evaluating the quality of an extracted policy is how accurately it matches the original policy. That means the authorization decisions made by the extracted policy for a set of access requests should be similar to the decisions made by the original policy for that set of requests. As an example, if the decision of the original policy for an access request $q$ is permit, then the decision of the mined policy for the same access request must be permit as well. If the mined policy denies the same access request, then we record this authorization tuple as a \textit{False Negative}. We define \textit{Relative True Positive}, \textit{Relative False Positive}, \textit{Relative True Negative}, and \textit{Relative False Negative} rates, respectively, as follows:

\begin{definition}
(\textbf{Relative True Positive Rate}).
Given an access log $\mathcal{L}$ and an ABAC policy $\pi$, the relative true positive rate of $\pi$ regarding $\mathcal{L}$ denoted as $TP_{\pi|\mathcal{L}}$ is the portion of positive access logs for which the decision of $\pi$  is \emph{permit}:

\begin{equation*}
TP_{\pi|\mathcal{L}} = \dfrac{|\{ \langle q,d \rangle \in \mathcal{L}^+ | d_\pi(q) = permit \}|}{|\mathcal{L}^+|}
\end{equation*}

Here, $|s|$ is the cardinality of set $s$.
\end{definition}

\begin{definition}
(\textbf{Relative False Positive Rate}).
The relative false positive rate of $\pi$ regarding $\mathcal{L}$ denoted as $FP_{\pi|\mathcal{L}}$ is the portion of negative access logs for which the decision of $\pi$ is \emph{permit}:

\begin{equation*}
FP_{\pi|\mathcal{L}} = \dfrac{|\{ \langle q,d \rangle \in \mathcal{L}^- | d_\pi(q) = permit \}|}{|\mathcal{L}^-|}
\end{equation*}
\end{definition}

Similarly, we calculate the relative true negative rate and false negative rate of $\pi$ regarding $\mathcal{L}$, denoted as $TN_{\pi|\mathcal{L}}$ and $FN_{\pi|\mathcal{L}}$, respectively, as follows:

\begin{equation*}
TN_{\pi|\mathcal{L}} = \dfrac{|\{\langle q,d \rangle \in \mathcal{L}^- | d_\pi(q) = deny \}|}{|\mathcal{L}^-|}
\end{equation*}

\begin{equation*}
FN_{\pi|\mathcal{L}} = \dfrac{|\{ \langle q,d \rangle \in \mathcal{L}^+ | d_\pi(q) = deny \}|}{|\mathcal{L}^+|}
\end{equation*}

The \textit{relative precision} and \textit{relative recall} are calculated as follows:

\begin{equation*}
Precision_{\pi|\mathcal{L}} = \dfrac{TP_{\pi|\mathcal{L}}}{TP_{\pi|\mathcal{L}} + FP_{\pi|\mathcal{L}}}
\end{equation*}

\begin{equation*}
Recall_{\pi|\mathcal{L}} = \dfrac{TP_{\pi|\mathcal{L}}}{TP_{\pi|\mathcal{L}} + FN_{\pi|\mathcal{L}}}
\end{equation*}

The relative accuracy metric, $ACC_{\pi|\mathcal{L}}$, measures the accuracy of mined policy $\pi$ with regards to the decisions made by the original policy indicated by $\mathcal{L}$ and is defined formally as follows:

\begin{definition}
(\textbf{Relative Accuracy}).
Given the relative true positive and negative rates, the relative accuracy of $\pi$ regarding $\mathcal{L}$ denoted as $ACC_{\pi|\mathcal{L}}$ is calculated as follows:

\begin{equation*}
ACC_{\pi|\mathcal{L}} = \dfrac{TP_{\pi|\mathcal{L}} + TN_{\pi|\mathcal{L}}}{TP_{\pi|\mathcal{L}} + TN_{\pi|\mathcal{L}} + FP_{\pi|\mathcal{L}} + FN_{\pi|\mathcal{L}}}
\end{equation*}
\end{definition}

As accuracy may be misleading in unbalanced data sets \cite{accuracy_paradox} (which is very probable in case of access logs), we use \textbf{relative F-score} to better evaluate the mined policy:

\begin{equation*}
F{\text -}score_{\pi|\mathcal{L}} = 2 \cdot \dfrac{Precision_{\pi|\mathcal{L}} \cdot Recall_{\pi|\mathcal{L}}}{Precision_{\pi|\mathcal{L}} + Recall_{\pi|\mathcal{L}}}
\end{equation*}

Policies with higher relative F-score are better as they are more consistent with the original access log.

On the other hand, as the number of filters in each rule and the number of rules in an access control policy increases, policy intelligibility would decrease and maintenance of the policy would become harder. Hence, complexity is another key metric for evaluating the quality of a policy. 

\textbf{Weighted Structural Complexity (WSC)} is a generalization of policy size and was first introduced for RBAC policies \cite{molloy2010mining} and later extended for ABAC policies \cite{xu2015mining}. WSC is consistent with usability studies of access control rules, which indicates that the more concise the policies are the more manageable they become  \cite{beckerle2013formal}. Informally, for a given ABAC policy, its WSC is a weighted sum of its elements. Formally, for an ABAC policy $\pi$ with rules $\mathcal{P}$, its WSC is defined as follows:
\begin{equation*}
  \begin{gathered}
  WSC(\pi) = WSC(\mathcal{P})
  \end{gathered}
\end{equation*}

\begin{equation*}
  \begin{gathered}
  WSC(\mathcal{P}) = \sum\limits_{\rho \in \mathcal{P}} WSC(\rho)
  \end{gathered}
\end{equation*}

\begin{equation*}
  \begin{gathered}
  WSC(\rho = \langle \mathcal{F_U}, \mathcal{F_O}, \mathcal{F_S}, \mathcal{R}, op, d \rangle ) = w_1 WSC(\mathcal{F_U}) + \\
  w_2 WSC(\mathcal{F_O})  + w_3 WSC(\mathcal{F_S}) + w_4 WSC(\mathcal{R})
  \end{gathered}
\end{equation*}

\begin{equation*}
  \begin{gathered}
  \forall s \in \{\mathcal{F_U}, \mathcal{F_O}, \mathcal{F_S}, \mathcal{R}\} : WSC(s) = \sum\limits |s|
  \end{gathered}
\end{equation*}

where $|s|$ is the cardinality of set $s$  and each $w_i$ is a user-specified weight.

Van Rijsbergen proposes an effectiveness measure for combining two different metrics $P$ and $R$ in \cite{rijsbergen_1979} as follows :

\begin{equation*}
  \begin{gathered}
  E = 1 - \dfrac{1}{\dfrac{\alpha}{P} + \dfrac{1 - \alpha}{R}}
  \end{gathered}
\end{equation*}

Given relative F-score and WSC measures for various mined policies resulting from running different mining algorithms over access log, it may not be straightforward to select the best algorithm and, hence, the mined policy with the highest quality. So, to be able to compare the quality of different mined ABAC policies, we combine the two metrics based on Van Rijsbergen's effectiveness measure \cite{rijsbergen_1979} and define the \textbf{Policy Quality Metric} as follows:

\begin{equation*}
  \begin{gathered}
  \mathcal{Q}_{\pi} = ( \dfrac{\alpha}{F{\text -}score_{\pi|\mathcal{L}}} +  \dfrac{1 - \alpha}{\Delta WSC_{\pi}})^{-1}
  \end{gathered}
\end{equation*}

Here $\alpha = \dfrac{1}{1 + \beta^2}$ where $\beta$ determines the importance of relative F-score over policy complexity and $\Delta WSC_{\pi}$ shows the relative reduction in the complexity with regards to the complexity of the most complex mined policy. $\Delta WSC_{\pi}$ is calculated as follows:
\begin{equation*}
  \begin{gathered}
  \Delta WSC_{\pi} = \dfrac{WSC_{max} - WSC(\pi) + 1}{WSC_{max}}
 \end{gathered}
\end{equation*} 
$WSC_{max}$ is the weighted structural complexity of the most complex mined policy. 

\begin{definition}
(\textbf{Most Complex Mined Policy}).
The most complex mined policy is the mined policy with the highest weighted structural complexity. It is extracted by iterating through positive access log $\mathcal{L^+}$ and adding an access control rule for each authorization tuple if it's not already included in the mined policy. The corresponding rule for each authorization tuple includes all attributes of user, object, and subject of that authorization tuple.
\end{definition}

Considering the equal importance of relative F-score and relative loss of complexity of the policy, we calculate the quality measure as follows:

\begin{equation*}
  \begin{gathered}
  \mathcal{Q}_{\pi} = \dfrac{2 \cdot F{\text -}score_{\pi|\mathcal{L}} \cdot \Delta WSC_{\pi}}{F{\text -}score_{\pi|\mathcal{L}} + \Delta WSC_{\pi}}
  \end{gathered}
\end{equation*}

A mined policy with a higher F-score would have a higher policy quality. On the other hand, as the complexity of a policy increases, its quality will decrease. The intuition here is that once an extracted policy reaches a high F-score, adding additional rules will lead to a decrease in $\mathcal{Q}_{\pi}$.

For the most complex mined policy $\pi_w$, $\Delta WSC_{\pi_w} \approx 0$, so its policy quality $\mathcal{Q}_{\pi_w}$ is very close to zero. For an empty mined policy $\pi_e$ (a policy without any rule), while $\Delta WSC_{\pi_e} \approx 1$, as it denies all the access requests, its false negative rate is one and its true positive rate is zero. So its precision is zero and as a result, its F-score is zero as well. So the quality of the empty policy $\mathcal{Q}_{\pi_e}$ is zero, too. 

The most complex mined policy and the empty mined policy are the two extreme cases with policy quality equal to zero. Other mined policies between these two cases have higher policy quality than zero. 

\section{The Proposed Learning-based Approach} \label{proposed}
\begin{figure}[htbp]
    \centering\includegraphics[scale=0.5]{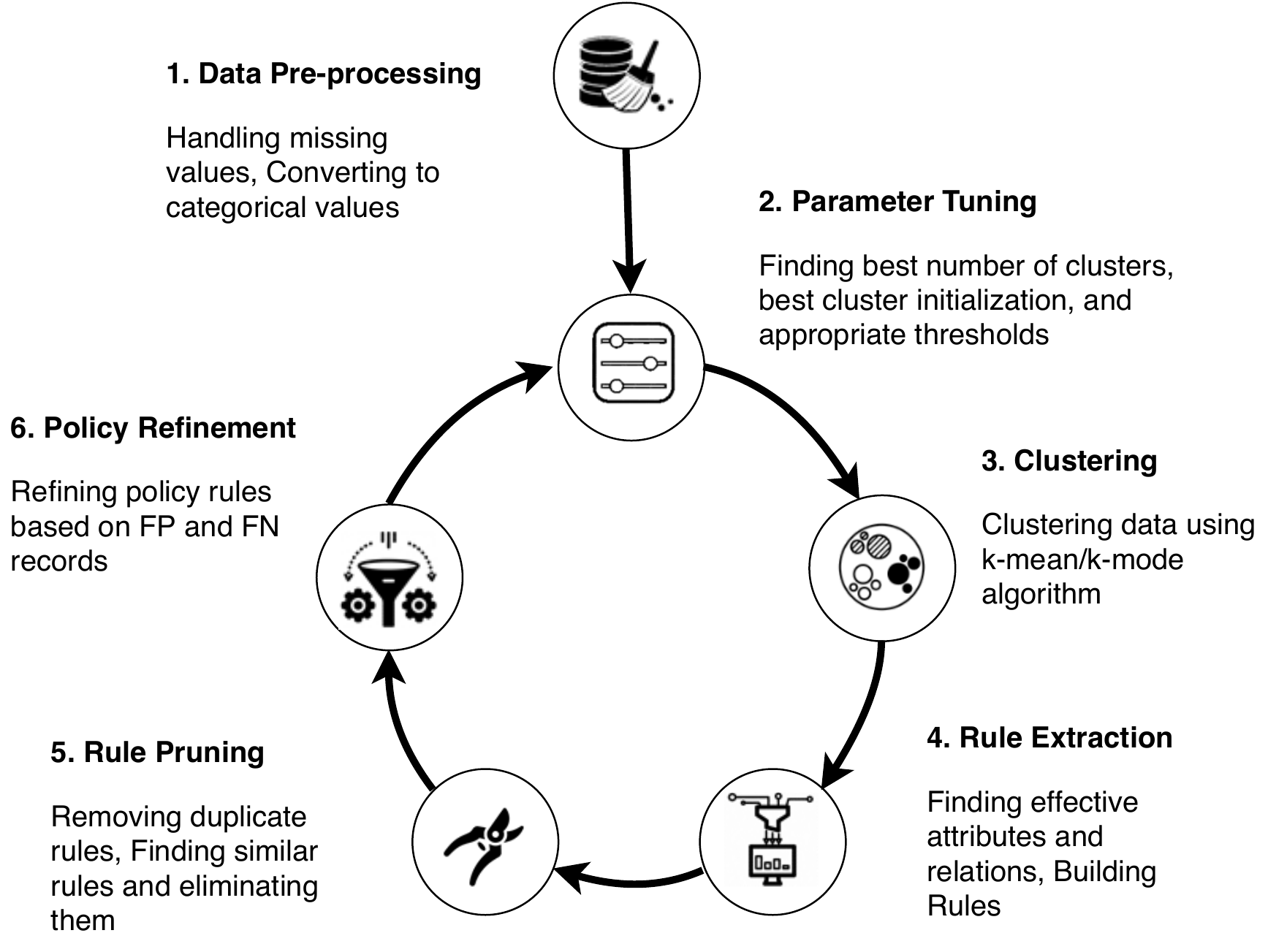}
    \caption{Overview of the Proposed Approach.}
    \label{fig:overview}
\end{figure}

Our proposed learning-based ABAC policy extraction pro-cedure  consists  of  the  steps  summarized  in  Figure  \ref{fig:overview}.

\subsection{Data Pre-processing}
As features of our learning algorithm are categorical variables, the first step in pre-processing the access log is to convert all numerical variables to their corresponding categorical values. For example, in ABAC, environmental attributes deal with time, location or dynamic aspects of the access control scenario. Hence, we need to pre-process and discretize such continuous variables to categorical ones (e.g. time of access to working hours and non working hours) so our proposed algorithm is applicable to them.

We also need to handle \emph{missing values} in this step. As the frequency of each attribute value is an important factor in our rule extraction algorithm (Section \ref{rule_extraction}) for deciding if an attribute is effective or not, it is important to replace missing values in a way that it doesn't mess up with the original frequency of each attribute value. For this purpose, we replace each missing value by \emph{UNK} (i.e., unknown).

\subsection{Selection of Learning Algorithm }
We use the \textit{K-modes algorithm} \cite{cao2009new}, which is a well known unsupervised learning algorithm used for clustering categorical data. \textit{K-modes} has been proved effective in mining ABAC policies \cite{karimi2018unsupervised}; this algorithm uses an initialization method based on both the distance between data points and the density of data points. Using both density and distance when initializing clusters help avoid two problems: (i) clustering outliers as new clusters are based only on the distances; and (ii) creating new clusters surrounding one center based only on the density. Compared to a random initialization method, this method provides more robustness and better accuracy in the clustering process\cite{cao2009new}.

\subsection{Parameter Tuning}
In the next step, we \emph{tune the learning parameters}. There are several challenges that need to be addressed in this step, which include the following: 

\subsubsection{Number of Clusters (k)}

One of the main challenges in an unsupervised learning is determining the number of clusters, $k$. In our sample policies, as we know the number of rules in each policy, we can set the number of clusters beforehand but in a real situation as we do not know the size of the rules in advance, making the correct choice of $k$ is difficult. One of the popular methods for determining the number of clusters in an unsupervised learning model is the \emph{Elbow Method} \cite{thorndike1953belongs,goutte1999clustering}. This method is based on total within group sum of squares. $k$ will be chosen as the number of clusters if adding another cluster doesn't give much better modeling of the data (i.e., the elbow point of the graph). 

As a second approach, we choose a number of clusters ($k$) which gives the best modeling of the data in terms of the policy \textit{quality} metric. For this purpose, we run our clustering algorithm for different values of $k$ and calculate the accuracy of the corresponding model using 10-fold cross-validation. The value of $k$ that maximizes the accuracy of the model is selected as the final number of clusters. 

Note that increasing $k$ will ultimately reduce the amount of clustering error or it will increase the accuracy of the model, but by increasing the number of clusters, the number of extracted rules will also increase  resulting in more complexity (i.e., higher $\mathit{WSC}$). So it is important to find an optimal $k$ that balances between policy accuracy and WSC.

\subsubsection{Cluster Initialization \& Local Optima}  Different cluster initializations can lead to a different set of clusters as \emph{k}-means/\emph{k}-modes may converge to a local optima. To overcome this issue, for a given number of clusters, $k$,  we train multiple models with different cluster initializations and then select the partition with the smallest clustering error. 

\subsection{Policy Rules Extraction} \label{rule_extraction}
The main phase in our proposed approach is the extraction of ABAC policy rules. In the first step, we need to collect all the authorization tuples related to each rule of the policy. We use data clustering for this purpose. We divide the access log into clusters where the records in each cluster correspond to one AC rule in the system. This is done based on finding similar patterns between features (i.e., attribute values) of the records (i.e., access control tuples). In the second step, we extract the \emph{attribute filters} of such a rule. We adapt the rule extraction algorithm in \cite{karimi2018unsupervised} and extend it to extract both positive and negative attribute filters. We define \emph{effective positive attribute} and \emph{effective negative attribute} as follows:

\begin{definition}
(\textbf{Effective Positive (Negative) Attribute}). Let $S =\{ \langle a, v \rangle \}$ be the set of all possible attribute-value pairs in a system; we define $\langle a_j, v_j \rangle \: \in S$ ($ \langle a_j, !v_j \rangle \in S$) as an \emph{effective positive (negative) attribute} pair of $\rho_i$ corresponding to cluster $C_i$, where the frequency of occurrence of $v_j$ in the set of all the records of cluster $C_i$ is much higher (lower) than its frequency of occurrence in the original data; this is determined based on a threshold $\mathcal{T}_P$ ($\mathcal{T}_N$). The attribute expression $ \langle a_j, v_j \rangle $ ($ \langle a_j, !v_j \rangle $) is added to the attribute filters of the extracted rule $\rho_i$ for $C_i$ .
\end{definition}

In the final step, we extract the \emph{relation conditions} for AC rules for each cluster. This will be done based on the frequency of equality between pairs of attributes in the records of each cluster. We define \emph{effective positive relation} and \emph{effective negative relation} as follows:

\begin{definition}
(\textbf{Effective Positive (Negative) Relation}). Let $R = \{ \langle a, b \rangle \}$ be the set of all possible relations between pairs of attributes in the system; we define $ \langle a_j, b_j \rangle $ as an \emph{effective positive (negative) relation} pairs of $\rho_i$ corresponding to cluster $C_i$, where the frequency of $a_j$ equals $b_j$ in all the records of cluster $C_i$ is much higher (lower) than their frequency in the original data; this is determined based on a threshold $\theta_P$ ($\theta_N$). The relation $ \langle a_j, b_j \rangle $ ($ \langle a_j, !b_j \rangle $) is added to the relation conditions of the extracted rule $\rho_i$ for this cluster.
\end{definition}

We note that the values of the thresholds $\mathcal{T}_P$, $\mathcal{T}_N$, $\theta_P$, and $\theta_N$ will be different for each data set. To find the best threshold values for each data set, we run the rule extraction algorithm for different values of thresholds, and the values which result in the maximum accuracy over the cross-validation data set will be selected.

Algorithms \ref{attrExtraction} and \ref{relExtraction} show effective attribute and effective relation extraction procedures, respectively. 

\begin{algorithm}
\caption{Effective attribute extraction algorithm}\label{attrExtraction}
\begin{algorithmic}[1]
\Procedure{extractAttributeFilters}{}
\Require {$C_i$, $A$, $V$, $\mathcal{L}$, $\mathcal{T}_P$, $\mathcal{T}_N$}
\Ensure {$\mathcal{F}$}
\State $\mathcal{F} \gets \oldemptyset$
\ForAll{$a \in A$}
\ForAll{$v_j \in V_{a}$}
\If {$Freq(v_j, C_i) - Freq(v_j, \mathcal{L}) > \mathcal{T}_P$}
\State $\mathcal{F}i \gets \mathcal{F} \: \cup \langle a, v_j \rangle $
\EndIf
\If {$Freq(v_j, \mathcal{L}) - Freq(v_j, C_i) > \mathcal{T}_N$}
\State $\mathcal{F}i \gets \mathcal{F} \: \cup \langle a, !v_j \rangle $
\EndIf
\EndFor
\EndFor

\Return $\rho_i$
\EndProcedure
\end{algorithmic}
\end{algorithm}

\begin{algorithm}
\caption{Effective relation extraction algorithm}\label{relExtraction}
\begin{algorithmic}[1]
\Procedure{extractRelations}{}
\Require {$C_i$, $A$, $\mathcal{L}$, $\theta_P$, $\theta_N$}
\Ensure {$\mathcal{R}$}
\State $\mathcal{R} \gets \oldemptyset$
\ForAll{$a \in A$}
\ForAll{$b \in A$ and $b \neq a$}
\If {$Freq(a= b, C_i)$ - $Freq(a=b, \mathcal{L})$>$\theta_P$}
\State $\mathcal{R} \gets \mathcal{R} \: \cup \langle a, b \rangle $
\EndIf
\If {$Freq(a= b, \mathcal{L})$ - $Freq(a=b, C_i)$>$\theta_N$}
\State $\mathcal{R} \gets \mathcal{R} \: \cup \langle a, !b \rangle $
\EndIf
\EndFor
\EndFor

\Return $\mathcal{R}$
\EndProcedure
\end{algorithmic}
\end{algorithm}

\subsection{Policy Enhancement}
After the first phase of policy rule extraction, we get a policy which may not be as accurate and concise as we desire. We enhance the quality of the mined policy through iterations of policy improvement steps that include: \emph{rule pruning} and \emph{policy refinement}. 
\subsubsection{Rule Pruning} \label{pruning}
During the rule extraction phase, it's possible to have two clusters that correspond to the same rule. As a result, the extracted rules of these clusters are very similar to each other. Having two similar rules in the final policy increases the complexity of the mined policy while it may not help the accuracy of the policy and as a result, it hurts the policy quality. To address such an issue, in the rule pruning step, we identify similar rules and eliminate the ones whose removal improves the policy quality more. If eliminating neither of the two rules improves the policy quality, we keep both the rules. This may happen when we have two very similar AC rules in the original policy. We measure the similarity between two rules using Jaccard similarity \cite{jaccard1912distribution} as follows:

\begin{equation*}
  \begin{gathered}
  J(S_1, S_2) = |S_1 \cap S_2| / |S_1 \cup S_2|
  \end{gathered}
\end{equation*}

Based on this, we calculate the similarity between two rules $\rho_1$ and $\rho_2$ as follows:
\begin{equation*}
  \begin{gathered}
  J(\rho_1, \rho_2) = \\
  \frac{\big[ \sum\limits_{\mathcal{F} \in \{\mathcal{F_U}, \mathcal{F_O}, \mathcal{F_S}\}} |\mathcal{F}_{\rho_1} \cap \mathcal{F}_{\rho_2}| + |\mathcal{R}_{\rho_1} \cap \mathcal{R}_{\rho_2}| + |op_{\rho_1} \cap op_{\rho_2}| \big]} {\big[ \sum\limits_{\mathcal{F} \in \{\mathcal{F_U}, \mathcal{F_O}, \mathcal{F_S}\}} |\mathcal{F}_{\rho_1} \cup \mathcal{F}_{\rho_2}| + |\mathcal{R}_{\rho_1} \cup \mathcal{R}_{\rho_2}| + |op_{\rho_1} \cup op_{\rho_2}| \big] }
  \end{gathered}
\end{equation*}

We consider two rules to be similar if their Jaccard similarity score is more than 0.5, which means that the size of their common elements is more than half of the size of the union of their elements. Algorithm \ref{rulePruning} shows the rule pruning procedure.

\begin{algorithm}
\caption{Rule Pruning algorithm}\label{rulePruning}
\begin{algorithmic}[1]
\Procedure{rulePruning}{}
 \Require {$\pi$}
 \Ensure {$\pi$}
 \State $\mathcal{P} \gets \pi.\mathcal{P}$
 \State $q \gets \Call{calcQuality}{\mathcal{P}}$
\ForAll{$\rho_i \in \mathcal{P}$}
\ForAll{$\rho_j \in \mathcal{P}$ and $\rho_i \neq \rho_j$}
\If {$\Call{Similarity}{\rho_i, \rho_j} > 0.5$}
    \State $\mathcal{P}_i \gets \mathcal{P}/\rho_i$
    \State $\mathcal{P}_j \gets \mathcal{P}/\rho_j$
    \State $q_i \gets \Call{calcQuality}{\mathcal{P}_i}$
    \State $q_j \gets \Call{calcQuality}{\mathcal{P}_j}$
    \If {$q_i >= q$ and $q_i >= q_j$}
    \State $\mathcal{P} \gets \mathcal{P}_i$
    \EndIf
    \If {$q_j >= q$ and $q_j >= q_i$}
    \State $\mathcal{P} \gets \mathcal{P}_j$
    \EndIf
\EndIf
\EndFor
\EndFor

\Return $\mathcal{P}$
\EndProcedure
\end{algorithmic}
\end{algorithm}

\subsubsection{Policy Refinement}
During the rule extraction phase, it is possible to extract rules that are either too restricted or too relaxed compared to the original policy rules. A rule is restricted if it employs more filters than the original rule.  

\begin{example} \label{exam6}
Consider the following two rules:
\begin{equation*}
    \begin{split}
        {\rho}_1 = \langle \{(position, faculty)\},& \{(type, gradebook)\}, \\
        \{setScore\}, permit \rangle \\
        {\rho}_2 = \langle \{(position, faculty),& (uDept, EE)\}, \\
        \{(type, gradebook)\}, &\{setScore\}, permit \rangle
    \end{split}
\end{equation*}
\end{example}

Here ${\rho}_2$ is more restricted than ${\rho}_1$ as it imposes more conditions on the user attributes.

Having such a restricted rule in the mined policy would result in a larger number of \emph{FNs} as an access request that  would  be  permitted by the  original  rule  will be denied by the restricted rule.

On the other hand, an extracted rule is more relaxed compared to the original rule if it misses some of the filters. In Example \ref{exam6}, ${\rho}_1$ is more relaxed than ${\rho}_2$. Such a relaxed rule would result in more \emph{FPs} as it permits access requests that should be denied as per the original policies.

To address these issues, we propose a \emph{policy refinement} procedure which is shown in Algorithm \ref{policyrefinement}. Here, we try to refine the mined policy ($\pi_m$) based on the patterns discovered in the FN or FP records. These patterns are used to eliminate extra filters from restricted rules or append missing filters to relax the rules. 

To extract patterns from the FN or FP records, we apply our rule extraction procedure on these records to get the corresponding policies $\pi_{FN}$ and $\pi_{FP}$. Here our training data are FN and FP records, respectively. We compare the extracted FN or FP rules with the mined policy and remove the extra filters or append the missed ones to the corresponding rules. As an example, consider the FP records. Here, our goal is to extract the patterns that are common between access requests that were permitted based on the mined policy while they should have been denied based on the original policy.

In each step  of  refinement,  a  rule  from $\pi_m$ that is similar to a rule from $\pi_{FN}$ or $\pi_{FP}$ based on the Jaccard similarity (Section \ref{pruning}) is selected and then refined in two ways as discussed below.

\vspace{2mm}
\noindent\textit{\textbf{Policy refinement based on $\pi_{FN}$}}:
In the case of FN records, two situations are possible: a rule is missing from the mined policy ($\pi_m$) or one of the rules in $\pi_m$ is more restrictive. To resolve this issue, for each rule $\rho_i \in \pi_{FN}$:

\begin{itemize}
    \item if there is a similar rule $\rho_j \in \pi_m$ then we refine $\rho_j$ as follows:
    
    \begin{equation*}
        \begin{gathered}
            \forall f \in \mathcal{F}  : {\mathcal{F}_{\rho}}_j = {\mathcal{F}_{\rho}}_j/({\mathcal{F}_{\rho}}_j/{\mathcal{F}_{\rho}}_i) \\
         \end{gathered}
    \end{equation*}
     where $\mathcal{F} = \mathcal{F_U} \cup \mathcal{F_O} \cup \mathcal{F_S} \cup \mathcal{R}$. So, the extra filters are removed from the restricted rule ($\rho_j$).
    
    \item if there is no such rule, then $\rho_i$ is the missing rule and we add it to $\pi_m$.
\end{itemize}

\noindent\textit{\textbf{Policy refinement based on $\pi_{FP}$}}:
In the case of FP records, some filters might be missing in an extracted rule in the mined policy ($\pi_m$); so for each rule $\rho_i \in \pi_{FP}$, we refine the mined policy as follows:
\begin{equation*}
        \begin{gathered}
            \forall f \in \mathcal{F}  : {\mathcal{F}_{\rho}}_j = {\mathcal{F}_{\rho}}_j \cup ({\mathcal{F}_{\rho}}_i/{\mathcal{F}_{\rho}}_j) \\
         \end{gathered}
    \end{equation*}
     where $\mathcal{F} = \mathcal{F_U} \cup \mathcal{F_O} \cup \mathcal{F_S} \cup \mathcal{R}$ includes all the filters in the rule. So, the missing filters are added to the relaxed rule ($\rho_j$).
     
These refinements can be done in multiple iterations until further refinement does not give a better model in terms of  policy quality $\mathcal{Q}_{\pi}$.

\begin{algorithm}
\caption{Policy refinement algorithm}\label{policyrefinement}
\begin{algorithmic}[1]
\Procedure{refinePolicy}{}
\Require {$A$, $\mathcal{L}$}
\Ensure {$\pi_m$}
\State $\mathcal{FN} \gets \Call{getFNs}{\pi_m, \mathcal{L}}$
\State $\pi_{FN} \gets \Call{extractPolicy}{\mathcal{FN}}$
\ForAll{$\rho_i \in \pi_{FN}.\mathcal{P}$}
\State $R_s \gets \Call{getSimilarRules}{\pi_{FN}.\mathcal{P}, \pi_m.\mathcal{P}}$
\If {$|R_s| = 0$}
\State $\pi_m.\mathcal{P} \gets \pi_m.\mathcal{P} \cup \rho_i$
\Else
\ForAll{$\rho_j \in R_s$}
    \ForAll{$\mathcal{F} \in \mathcal{F_U} \cup \mathcal{F_O} \cup \mathcal{F_S} \cup \mathcal{R}$}
        \State $\mathcal{F}_{\rho_j} \gets \mathcal{F}_{\rho_j} \backslash (\mathcal{F}_{\rho_j}\backslash \mathcal{F}_{\rho_i})$
    \EndFor
\EndFor
\EndIf
\EndFor

\State $\mathcal{FP} \gets \Call{getFPs}{\pi_m, \mathcal{L}}$
\State $\pi_{FP} \gets \Call{extractPolicy}{\mathcal{FP}}$
\ForAll{$\rho_i \in \pi_{FP}.\mathcal{P}$}
\State $R_s \gets \Call{getSimilarRules}{\pi_{FP}.\mathcal{P}, \pi_m.\mathcal{P}}$
\If {$|R_s| \: != 0$}
\ForAll{$\rho_j \in R_s$}
    \ForAll{$\mathcal{F} \in \mathcal{F_U} \cup \mathcal{F_O} \cup \mathcal{F_S} \cup \mathcal{R}$}
        \State $\mathcal{F}_{\rho_j} \gets \mathcal{F}_{\rho_j} \cup (\mathcal{F}_{\rho_i}\backslash \mathcal{F}_{\rho_j})$
    \EndFor
\EndFor
\EndIf
\EndFor

\Return $\pi_m$
\EndProcedure
\end{algorithmic}
\end{algorithm}

\section{Experimental Evaluation} \label{evaluation}
We have implemented a  prototype  of  our proposed approach presented in Section \ref{proposed}. Here, we present our experimental evaluation.

\subsection{Datasets}
We perform our experiments on multiple datasets including synthesized and real ones. The synthesized access logs are generated from two sets of ABAC policies. The first one is a manually written set of policies that is adapted from \cite{xu2015mining} to be compatible with our policy language. The second one includes a completely randomly generated set of policies. To synthesize our input data, for each ABAC policy (i.e., \emph{University Policy}, \emph{Healthcare Policy}, etc.), a set of authorization tuples is generated and the outcome of the ABAC policy for each access right is evaluated. The authorization tuples with \emph{permit} as their outcomes are the inputs to our unsupervised learning model.

\begin{table*}
\centering
\caption{Details of the Synthesized and Real Policies} \label{tab:policies_details} 
  \begin{tabular}{clcccccc}
    \toprule
    $\#$ & $\pi$ & $|\mathcal{P}|$ & $|A|$ & $|V|$ & $|\mathcal{L}|$ & $|\mathcal{L}^+|$ & $|\mathcal{L}^-|$ \\
    \midrule
    $\pi_1$ & UniversityP & 10 & 11 & 45 & 2,700K & 231K & 2,468K \\
    $\pi_2$ & HealthcareP & 9 & 13 & 40 & 982K & 229K & 753K \\
    $\pi_3$ & ProjectManagementP & 11 & 14 & 44 & 5,900K & 505K &5,373K\\
    $\pi_4$ & UniversityPN & 10 & 11 & 45 & 2,700K & 735K & 1,964K \\
    $\pi_5$ & HealthcarePN & 9 & 13 & 40 & 982K & 269K & 713K \\
    $\pi_6$ & ProjectManagementPN & 11 & 14 &44 & 5,900K & 960K & 4,918K\\
    $\pi_7$ & Random Policy 1 & 10 & 8 & 27 & 17K &2,742 & 14K \\ 
    $\pi_8$ & Random Policy 2 & 10 & 10 & 48 & 5,250K & 245K & 5,004K \\
    $\pi_9$ & Random Policy 3 & 10 & 12 & 38 & 560K &100K &459K\\
    $\pi_{10}$ & Amazon Kaggle & - & 10 & 15K & 32K & 30K & 1897 \\ 
    $\pi_{11}$ & Amazon UCI & - & 14 & 7,153 & 70K & 36K & 34K\\
    \bottomrule
  \end{tabular}
\end{table*}

Our real datasets are built from access logs provided by Amazon in Kaggle competition \cite{kaggle} and available in the UCI machine learning repository \cite{uci_amazon_access}.

\textbf{Manual Policy - University:}
This policy is adapted from \cite{xu2015mining} and it controls access of different users including students, instructors, teaching assistants, etc., to various objects (applications, gradebooks, etc.). 

\textbf{Manual Policy - Healthcare:}
This  policy is adapted from \cite{xu2015mining} and is used to control access by different users (e.g. nurses, doctors, etc.) to electronic health records (EHRs) and EHR items.

\textbf{Manual Policy - Project Management:}
This  policy is adapted from \cite{xu2015mining} and it controls access by different users (e.g. department managers, project leaders,  employees,  etc.)  to  various  objects  (e.g.  budgets,schedules and tasks).

\textbf{Random Policies:}
The authorization rules for this policy is generated completely randomly from random sets of attributes and attribute values. These randomly generated policies provide an opportunity to evaluate our proposed algorithm on access logs with various sizes and with varying structural characteristics. However, we note that, the performance of our algorithm on random policies might not be representative of its performance in real scenarios and over real policies.

\textbf{Real Dataset - Amazon Kaggle:}
The Kaggle competition dataset \cite{kaggle} includes access requests made by Amazon’s
employees over two years. Each record in this dataset describes an employee’s request to a resource and whether the request was authorized or not. A record consists of the employee’s attribute values and the resource identifier. The dataset includes more than 12,000 users and 7,000 resources.   

\textbf{Real Dataset - Amazon UCI:}
This dataset is provided by Amazon in the UCI machine learning repository \cite{uci_amazon_access}. It includes more than 36,000 users and 27,000 permissions. Since the dataset contains over 33,000 attributes, our focus in this experiment is narrowed only to the most requested 8 permissions in the dataset. 

\textbf{Partial Datasets:} To check the efficiency of the proposed algorithm over sparse datasets, we generate sparse datasets (partial datasets) by randomly selecting authorization tuples from the complete dataset. For example, a 10\% sparse (partial) dataset is generated by randomly selecting 10\% of tuples from the complete access logs.

\textbf{Noisy Datasets:} To check the efficiency of the proposed algorithm over noisy datasets, we generate noisy datasets by randomly reversing the decision of authorization tuples. For instance, a 10\% noisy dataset is generated by randomly reversing the decision of 10\% of authorization tuples in the complete access logs. 

For each of the manual policies, we consider two different sets of policy rules; the first one only contains positive attribute filters and relations while the second one includes both positive and negative attribute filters and relations. We have included these policies in Appendix A.

Table \ref{tab:policies_details} shows the details of the manual and random access log datasets. In this table, $|\mathcal{P}|$ shows the number of rules in the original policy, $|A|$ and $|V|$ show the number of attributes and attribute values and $|\mathcal{L}|$, $|\mathcal{L}^+|$, $|\mathcal{L}^-|$ show the number of access control tuples, the number of positive access logs, and the number of negative access logs in the given dataset, respectively.

\subsection{Experimental Setup}
To evaluate our proposed method, we use a computer with 2.6 GHz Intel Core i7 and 16 GB of RAM. We use Python 3 in the mining and the evaluation process. The algorithms were highly time-efficient (e.g., maximum time consumption is less than half an hour).

We use kmodes library \cite{kmodes_implementation} for clustering our data. The initialization based on density (CAO) \cite{cao2009new} is chosen for cluster initialization in kmodes algorithm. 

To find optimal $k$, we apply the Silhouette method to test different values of $k$. We examine each value of $k$ in pre-defined set [10, 20]. Then the $k$ value that results in the highest Silhouette score is used in the final model.

To generate the synthesized access log $\mathcal{L}$, we brute force through all attributes $A$  and their values $V_{a}$ to produce all possible combinations for the tuples. This method was used to generate a complete access log for the random and manual policy datasets. We generate two sets of partial datasets; the 10\% partial datasets are used to check the efficiency of the proposed approach over sparse datasets (Table \ref{tab:our_mined_policies}) and the 0.1\% partial datasets are used to compare the proposed approach with previous work (Table \ref{tab:policies_comparison}). We also generate a set of noisy datasets to check the efficiency of the proposed algorithm over noisy access log. The results of such experiments are reported in Table \ref{tab:our_mined_policies}. 

For all experiments, the optimal thresholds for selecting effective attributes and relations are between 0.2 and 0.3.




\subsection{Results}\label{result}
We first evaluate the performance of our policy mining algorithm on complete datasets. Table \ref{tab:our_mined_policies} shows the results of these experiments.

Our second set of experiments is on partial datasets. The algorithm proposed by Xu and Stoller \cite{xu2014mining} and the approach presented by Cotrini \textit{et al.} \cite{cotrini2018mining} are not able to handle complete datasets as these datasets are huge. To be able to compare the performance of our proposed algorithm with their work, we generated 0.1\% sparse (partial) datasets and run all algorithms over these partial datasets. The results of these experiments are shown in Table \ref{tab:policies_comparison} and Figures \ref{fig:fscore_comparison},  \ref{fig:wsc_comparison}, and  \ref{fig:quality_comparison}.

The algorithm proposed by Xu and Stoller \cite{xu2014mining} and the approach presented by Cotrini \textit{et al.} \cite{cotrini2018mining} do not generate policy rules with negative attribute filters and relations, however we report the results of their algorithms over  datasets related to policy rules including negations (policies $\pi_4$, $\pi_5$, $\pi_6$) to show how the quality of mined policies would be impacted if the mining algorithm does not extract rules that include negation.

\subsubsection{The F-Score of the Mined Policies}
 Table \ref{tab:our_mined_policies} shows the final $F{\text -}score_{\pi|\mathcal{L}}$ of our proposed approach after several rounds of refinement over all complete datasets. As we can see in Table \ref{tab:our_mined_policies}, the proposed approach achieves high F-score across all experiments except for $\pi_6$. $\pi_6$ is a very complex dataset with both positive and negative attributes and relation filters including 14 attributes, 44 attribute values, and around six million access records. The final policy quality for this dataset is around 0.63, which is acceptable considering the complexity of the policy.
 
\begin{figure}[htbp]
    \centering\includegraphics[scale=0.4]{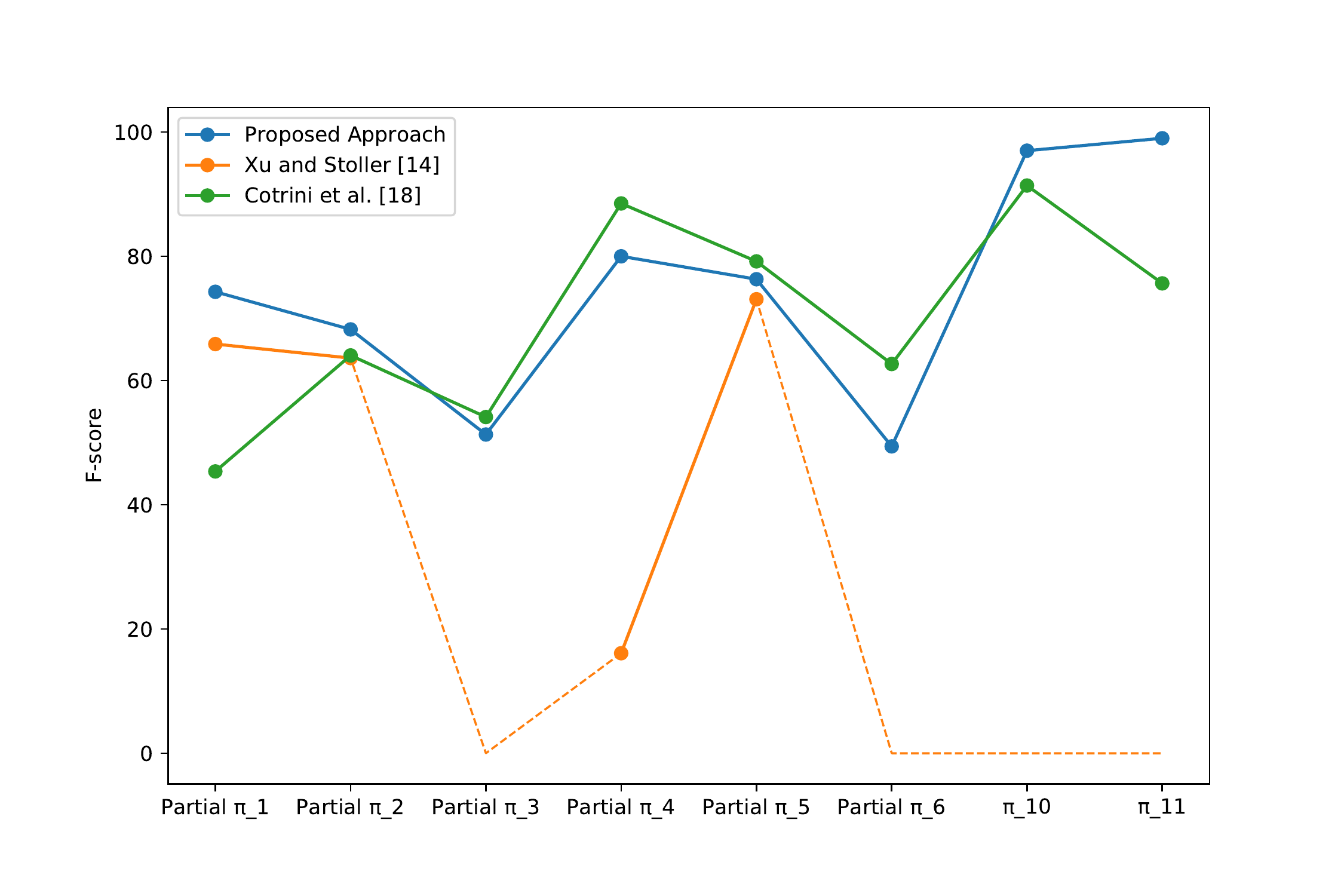}
    \caption{The F-Score of the Policies Mined by ABAC Mining Algorithms}
    \label{fig:fscore_comparison}
\end{figure}

 Table \ref{tab:policies_comparison} and Figure \ref{fig:fscore_comparison} show the comparison of the F-Scores of policies mined by our proposed approach with that of previous work over partial datasets (with 0.1\% of the complete datasets). The F-Score of policies mined by our algorithm is very close to the one done by the approach proposed by Cotrini \textit{et al.} \cite{cotrini2018mining}. As we can see, our proposed approach outperforms theirs in half of the experiments.
 
 
 \subsubsection{The Complexity of the Mined Policies}
 In Table \ref{tab:our_mined_policies}, we can see the final $WSC$ of the policies mined by our proposed approach. All extracted policies have the complexity lower than 100 which is much lower than those of the most complex policies for individual datasets. According to \textit{Definition 17}, the most complex policy for each dataset has the same complexity as the original positive access log ($\mathcal{L^+}$). Given numbers in Tables \ref{tab:policies_details} and \ref{tab:our_mined_policies}, the most complex policies for these scenarios are thousands of times more complex than the extracted policies by our approach.
 
\begin{figure}[htbp]
    \centering\includegraphics[scale=0.4]{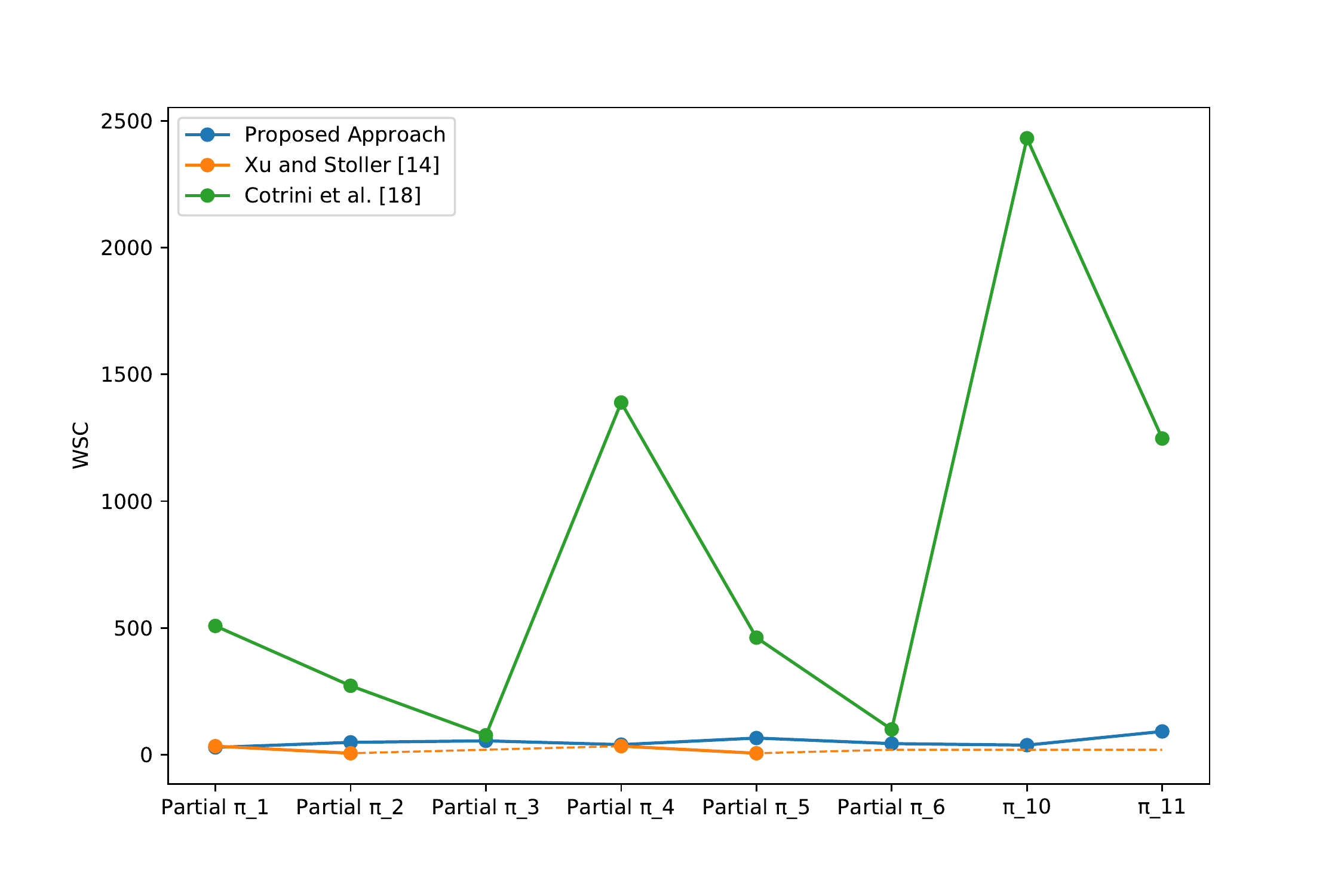}
    \caption{The Complexity of the Policies Mined by ABAC Mining Algorithms}
    \label{fig:wsc_comparison}
\end{figure}

 We compare the complexity of the policies mined by different ABAC mining algorithms in Figure \ref{fig:wsc_comparison}. Among three different approaches, the Cotrini \textit{et al.} algorithms extracts the most complex policies with WSC greater than 1000 for some cases. The complexity of the policies mined by our algorithm is very close to the one extracted by the approach proposed by Xu and Stroller \cite{xu2014mining}.
 
 \subsubsection{The Policy Quality of the Mined Policies}
 Finally, Table \ref{tab:our_mined_policies} shows the quality of the extracted policies through our proposed approach. We can see that out of all datasets that our proposed algorithm was applied on, around 75\% of the cases reached the policy quality of more than 0.8, which is significant, considering the huge size of original access logs (each more than 30K records).
 
\begin{figure}[htbp]
    \centering\includegraphics[scale=0.4]{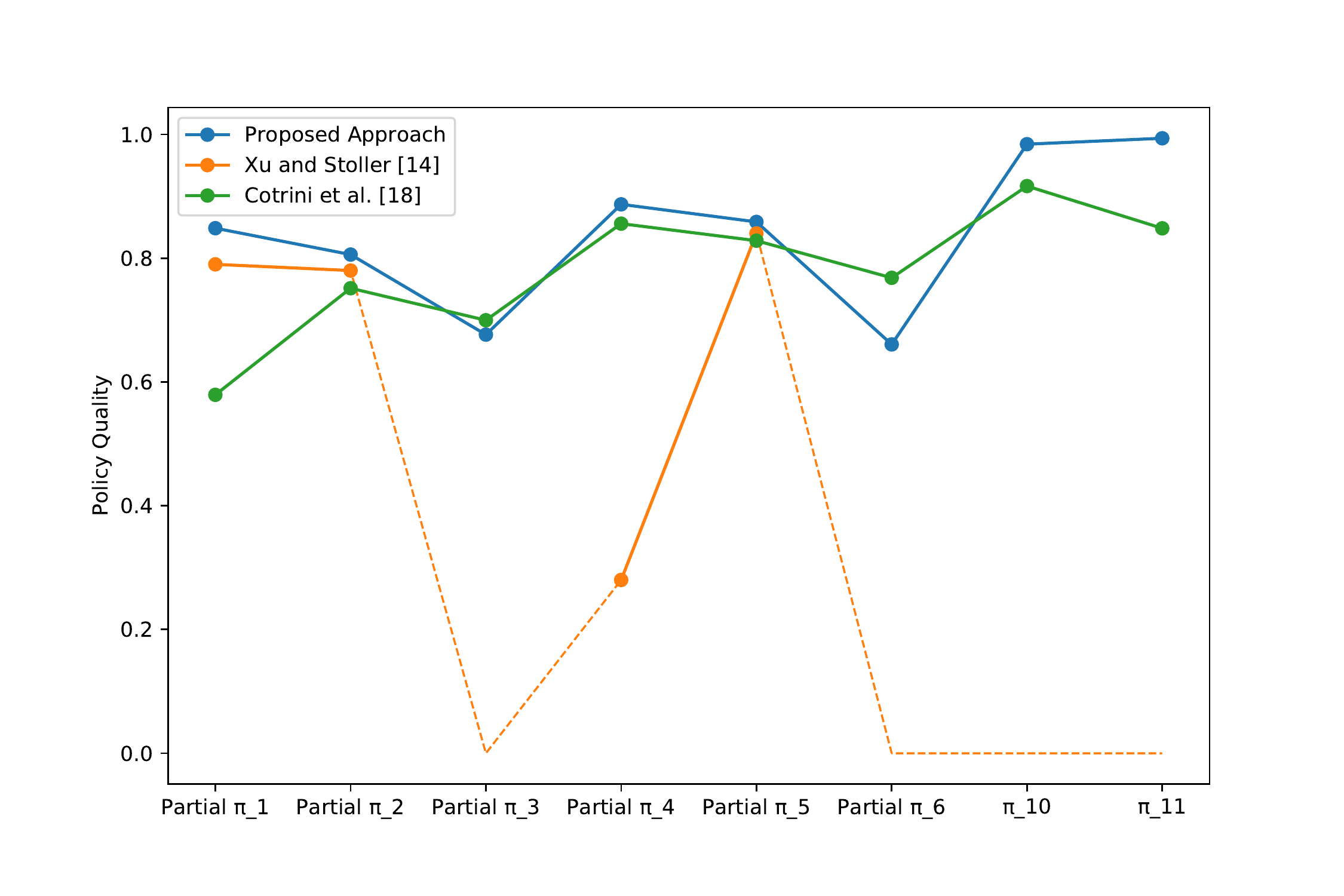}
    \caption{The Quality of the Policies Mined by ABAC Mining Algorithms}
    \label{fig:quality_comparison}
\end{figure}

 According to Figure \ref{fig:quality_comparison}, in most cases the policy quality of the policies mined by our proposed approach is higher than those of the policies extracted by other ABAC mining algorithms. 
 
\begin{table*}
\centering
\caption{Results of Our Proposed Approach on Various Synthesized and Real Policy Datasets} \label{tab:our_mined_policies} 
  \begin{tabular}{ccccccccc}
    \toprule
    $\pi$ & Total Running Time (s) & Optimal $k$ & $\mathcal{P}_{mined}$ & $ACC_{\pi|\mathcal{L}}$ & $F{\text -}score_{\pi|\mathcal{L}}$ & $WSC_{orig}$ & $WSC_{mined}$ & $\mathcal{Q}_{\pi}$ \\
    \midrule
    $\pi_1$ & 9376.556  & 15 & 20 & 97.5\% & 83.6\% & 33 & 91 & 0.91 \\
    Partial $\pi_1$ (10\%) & 1994.769  & 15 & 13 & 97.29\% & 82.21\% & 33 & 54 & 0.90\\
    Noisy $\pi_1$ (10\%) & 4979.56 & 10 & 8 & 96.94\% & 80\% & 33 & 28  & 0.90 \\
    $\pi_2$ & 2180.745 & 18 & 18& 85.49\% & 75.93\% & 33 & 71 & 0.86 \\
    Partial $\pi_2$ (10\%) & 4787.98 & 10 & 8 & 96.94\% &85.33\% & 33 & 28 & 0.92 \\
    Noisy $\pi_2$ (10\%) & 7339.91 &8 & 15 & 72.22\% & 82.13\% & 33 & 27 & 0.90 \\
    $\pi_3$ & 7795.44 &15 & 17&95.6\% &65.63\% & 44 & 55& 0.80\\
    Partial $\pi_3$ (10\%) & 1347.29 & 6 & 10 & 95.2\% &62.24\% & 44 & 56 & 0.77 \\
    Noisy $\pi_3$ (10\%) & 1912.72 & 15 & 15 & 94.47\% &62.66\% & 44 & 81 & 0.77\\
    $\pi_4$ & 13662.62 & 7 &16 &86.7\% & 71.58\% & 33 & 40 &0.83\\
    $\pi_5$ & 8681.64 & 15&15 & 78.11\% &62\% & 33 & 67&0.76\\
    $\pi_6$ & 12905.78 & 20&17 &88.05\% &46.28\% & 44 & 80 & 0.63 \\
    $\pi_7$ & 24.63 & 8 & 20 & 93\% &78.33\% &33 & 65 & 0.88\\
    $\pi_8$ & 13081.20 & 10 & 14 & 99.12\% & 91.28\% & 33& 51 & 0.95\\
    $\pi_9$ & 2266.68 & 8 & 16 & 92.17\% & 79.66\% &33 & 46 & 0.89\\
    $\pi_{10}$ & 265.3 & 15 & 20 & 94\% & 97\% & - & 44 & 0.98\\
    $\pi_{11}$ & 1010.43 & 24 & 25 & 98.49\% &99\% & - & 92 & 0.82\\
    \bottomrule
  \end{tabular}
\end{table*}

\begin{table*}
\centering
\begin{threeparttable}
\caption{Comparison of Our Proposed Approach with Previous Work on Various Synthesized and Real Policy Datasets} \label{tab:policies_comparison} 
  \begin{tabular}{lccccccc}
    \toprule
    Mining Alg. & $\pi$ & Time (s) & $ACC_{\pi|\mathcal{L}}$ & $F{\text -}score_{\pi|\mathcal{L}}$ & $\mathcal{P}_{\pi_{mined}}$ & $WSC(\pi)$ & $\mathcal{Q}_{\pi}$ \\
    \midrule
    Xu and Stoller \cite{xu2014mining} & \multirow{2}{*}{Partial $\pi_1$ (0.1\%)} & 227 & 94.74\% & 65.87\% & 10 & 34 & 0.79\\
    Cotrini \textit{et al.} \cite{cotrini2018mining} & &126 & 80.74\% & 45.3\% & 132 & 508 & 0.58 \\
    Proposed Approch & & 7.3 & 96\% & 74.2\% & 7 &29 & 0.85\\
    \hline
    Xu and Stoller \cite{xu2014mining} & \multirow{2}{*}{Partial $\pi_2$ (0.1\%)} & 32645 & 64.43 & 63.61 & 3 & 6 & 0.78 \\
    Cotrini \textit{et al.} \cite{cotrini2018mining} & &529 & 72.72\% & 64\% & 65 & 272 & 0.75\\
    Proposed Approch & & 7.9 & 79.78\% & 68.23\% & 13 & 49 & 0.81 \\
    \hline
    Xu and Stoller \cite{xu2014mining} & \multirow{2}{*}{Partial $\pi_3$ (0.1\%)} & $-^*$ & $-^*$ & $-^*$ & $-^*$ & $-^*$ & $-^*$ \\
    Cotrini \textit{et al.} \cite{cotrini2018mining} & & 3587 & 91.57\%& 54.124\%&24 &77 & 0.70\\
    Proposed Approch & & 11.44 & 94.96\%&51.31\% &12 &55 &0.78\\
    \hline
    Xu and Stoller \cite{xu2014mining} & \multirow{2}{*}{Partial $\pi_4$ (0.1\%)} & 4230 & 73.37\% & 16.1\% & 10 & 34 & 0.28 \\
    Cotrini \textit{et al.} \cite{cotrini2018mining} & & 204 & 93.55\%& 88.5\% &385 & 1389 & 0.86\\
    Proposed Approch & & 15 & 89.3\% & 80\% & 10 & 40 & 0.89\\
    \hline
    Xu and Stoller \cite{xu2014mining} & \multirow{2}{*}{Partial $\pi_5$ (0.1\%)} & 45348 & 79.25 & 73.09 & 3 & 6 & 0.84 \\
    Cotrini \textit{et al.} \cite{cotrini2018mining} & & 3587 & 86.46\%& 79.2\% & 123 &462 & 0.83 \\
    Proposed Approch && 8.8 & 87.2\% & 76.3\% & 15 & 66& 0.86\\
    \hline
    Xu and Stoller \cite{xu2014mining} & \multirow{2}{*}{Partial $\pi_6$ (0.1\%)} & $-^*$ & $-^*$ & $-^*$ & $-^*$ & $-^*$ & $-^*$ \\
    Cotrini \textit{et al.} \cite{cotrini2018mining} & & 2848 & 82.75\% & 62.66\% &31 &100 & 0.77\\
    Proposed Approch & &22.67 &81.2\% &49.4\% &12 &44 & 0.66\\
    \hline
    Xu and Stoller \cite{xu2014mining} & \multirow{2}{*}{$\pi_{10}$} & $-^*$ & $-^*$ & $-^*$ & $-^*$ & $-^*$ & $-^*$ \\
    Cotrini \textit{et al.} \cite{cotrini2018mining} & & 237  &84.25\% & 91.39\%&1055 &2431 &0.92 \\
    Proposed Approch & & 265.3 & 94\% & 97\% & 20 & 44 &0.98\\ \hline
    Xu and Stoller \cite{xu2014mining} & \multirow{2}{*}{$\pi_{11}$} & $-^*$ & $-^*$ & $-^*$ & $-^*$ & $-^*$ & $-^*$ \\
    Cotrini \textit{et al.} \cite{cotrini2018mining} & & 1345 & 70.93\%&75.64\% & 466 & 1247 &0.85 \\
    Proposed Approch & & 1010.43 & 98.49\% & 99\% & 24 & 92 &0.99\\
    \bottomrule
  \end{tabular}
  \begin{tablenotes}
    \item  $*$ Xu and Stoller \cite{xu2014mining} did not terminate nor produced any output for the these datasets even after running for more than 24 hours.
  \end{tablenotes}
  \end{threeparttable}
\end{table*}

\section{Related Work} \label{relatedwork}
As RBAC approach became popular, many organization decided to equip their information systems with more recent access control model, however migrating to RBAC from legacy access control systems was a huge obstacle for such environments. As a result, several researchers have addressed such a challenge by introducing automated role extraction algorithms \cite{molloy2010mining, xu2012algorithms, kuhlmann2003role, schlegelmilch2005role, vaidya2007role, vaidya2006roleminer, zhang2007role, guo2008role, molloy2008mining, takabi2010stateminer, ni2009automating}. Role engineering or role mining are the terms that have been used to refer to procedures to extract an optimal set of roles given user-permission assignments. 

In \cite{kuhlmann2003role}, Kuhlmann and Schimpf try to discover a set of roles from user-permission assignments using clustering techniques, however, they do not show the feasibility of their proposed approach through experiments. In addition, their proposed approach lacks a metric to choose the best model based on their clustering method. 

The ORCA role mining tool is proposed by Schlegelmilch and Steffens and tries to perform a hierarchical clustering on user-permission assignments \cite{schlegelmilch2005role}. Their proposed method limits the hierarchical structure to a tree so that each permission/user is assigned to one role in the hierarchy. This feature limits the feasibility of their proposed approach as, in real environments, roles do not necessarily form a tree.

Ni et al. propose a supervised learning approach for role mining which maps each user-permission assignment to a role using a supervised classifier (i.e., a support vector machine (SVM)) \cite{ni2009automating}. The main limitation of their proposed approach is that the roles and some parts of the role-permission assignments are needed beforehand; and hence, it is not applicable in many organizations. 

Vaidya \textit{et al.} are the first to define the Role Mining Problem (RMP) formally and analyze its
theoretical bounds \cite{vaidya2010role}. They also propose a heuristic approach for finding a minimal set of roles for a given set of user-permission assignments.

Xu and Stoller are the first to propose an algorithm for mining ABAC policies from RBAC \cite{xu2013mining}, logs \cite{xu2014mining}, and access control list \cite{xu2015mining} plus attribute information. Their policy mining algorithms iterate over access control tuples (generated from available information, e.g., user permission relations and attributes) and construct candidates rules. They then generalize the candidate rules by replacing conjuncts in attribute expressions with constraints. The main limitation of these algorithms is that as they are based on heuristic approaches, the proposed techniques work very well for simple and small scale AC policies, however, as the number of rules in the policy and the number of elements in each rule increases, they do not perform well.

Following Xu and Stroller's proposed method, Medvet \textit{et al.} \cite{medvet2015evolutionary} propose a multi-objective evolutionary algorithm for extracting ABAC policies. The proposed approach is a separate and conquer algorithm, in each iteration of which, a new rule is learned and the set of access log tuples becomes smaller. Their algorithm employs several search-optimizing features to improve the quality of the mined rules. Although their approach is a multi-objective optimization framework which incorporates requirements on both correctness and expressiveness, it suffers from the same issue as \cite{xu2015mining}.

Iyer and Masoumzadeh \cite{iyer2018mining} propose a more systematic, yet heuristic ABAC policy mining approach which is based on the rule mining algorithm called PRISM. It inherits shortcomings associated with PRISM that includes dealing with a large dimensionality of the search space of attribute values and generation of a huge number of rules. 

Cotrini \textit{et al.} propose an algorithm called Rhapsody for mining ABAC rules from sparse logs \cite{cotrini2018mining}. Their proposed approach is built upon subgroup discovery algorithms. They define a novel metric, \textit{reliability} which measures how overly permissive an extracted rule is. In addition, they propose a universal cross-validation metric for evaluating the mined policy when the input log is sparse. However, their algorithm is not capable of mining policies from logs with many attributes as the number of extracted rules grows exponentially in the number of attributes of the system.

\section{Discussion and Limitations}
As mentioned in section \ref{result}, our proposed approach is able to achieves a practical level of performance when applied to both synthesized and real datasets. In the case of synthesized datasets, the proposed approach is capable of mining policies containing both positive and negative attribute filters from complete datasets. On the other hand, our proposed approach shows potential for use in sparse datasets. In addition, the real datasets contain a large number of attributes and attribute values as shown in Table \ref{tab:policies_details}. The ability of our proposed approach in mining high-quality policies for these datasets shows that the size of attributes and attribute values have minimal impact on the effectiveness of our approach. 


 
The proposed approach is based on an unsupervised clustering algorithm. Since finding the proper number of clusters is a challenge related to clustering algorithms, our approach is affected by this issue as well. The same issue will also be valid in finding the best thresholds to extract effective attributes and relations.

We note that, as the proposed algorithm is based on tuning multiple parameters, it is possible that it gets stuck in minimum optima. For this reason, we do not claim that it will extract the policy with the highest quality in every scenario, nor we claim that extracting rules with negative attribute filters and relations would always result in policy with higher quality (as we can see in Section \ref{result}); however, by trying more randomization in cluster initialization and a wider range of parameters, we can get one that is closer to global optima.

In our evaluation, we used random selection to create noisy and sparse datasets from complete datasets. Although we ensured the same percentage of randomly selected tuples from permitted and denied logs, guaranteeing the quality of the sampling is difficult.

\section{Conclusion} \label{conclusion}
In this paper, we have proposed an unsupervised learning based approach to automating an ABAC policy extraction process. The proposed approach is capable of discovering both positive and negative attribute expressions as well as positive and negative relation conditions while previous approaches in access control policy extraction had only focused on positive expressions. Furthermore, our work is capable of improving the extracted policy through iterations of proposed rule pruning and policy refinement algorithms. Such refinement algorithms are based on the false positive and false negative records and they help in increasing the quality of the mined policy. 

Most importantly, we have proposed the \emph{policy quality metric} which considers both the conciseness and correctness of the mined policy and is important for comparing the extracted policy with the original one and for improving it as needed.

We have evaluated our policy extraction algorithm on access logs generated for various sample policies and demonstrated its feasibility. Furthermore, we have shown that our approach outperforms previous works in terms of policy quality.

As future work, we plan to extend our method to support numerical data and extract negative authorization rules as well while studying the effects of various conflict resolution strategies on the quality of the mined policy.


\ifCLASSOPTIONcaptionsoff
  \newpage
\fi



\bibliographystyle{ieeetr}
\bibliography{bibliography}


%





%



\begin{IEEEbiography}
    [{\includegraphics[width=1in,height=1.25in,clip,keepaspectratio]{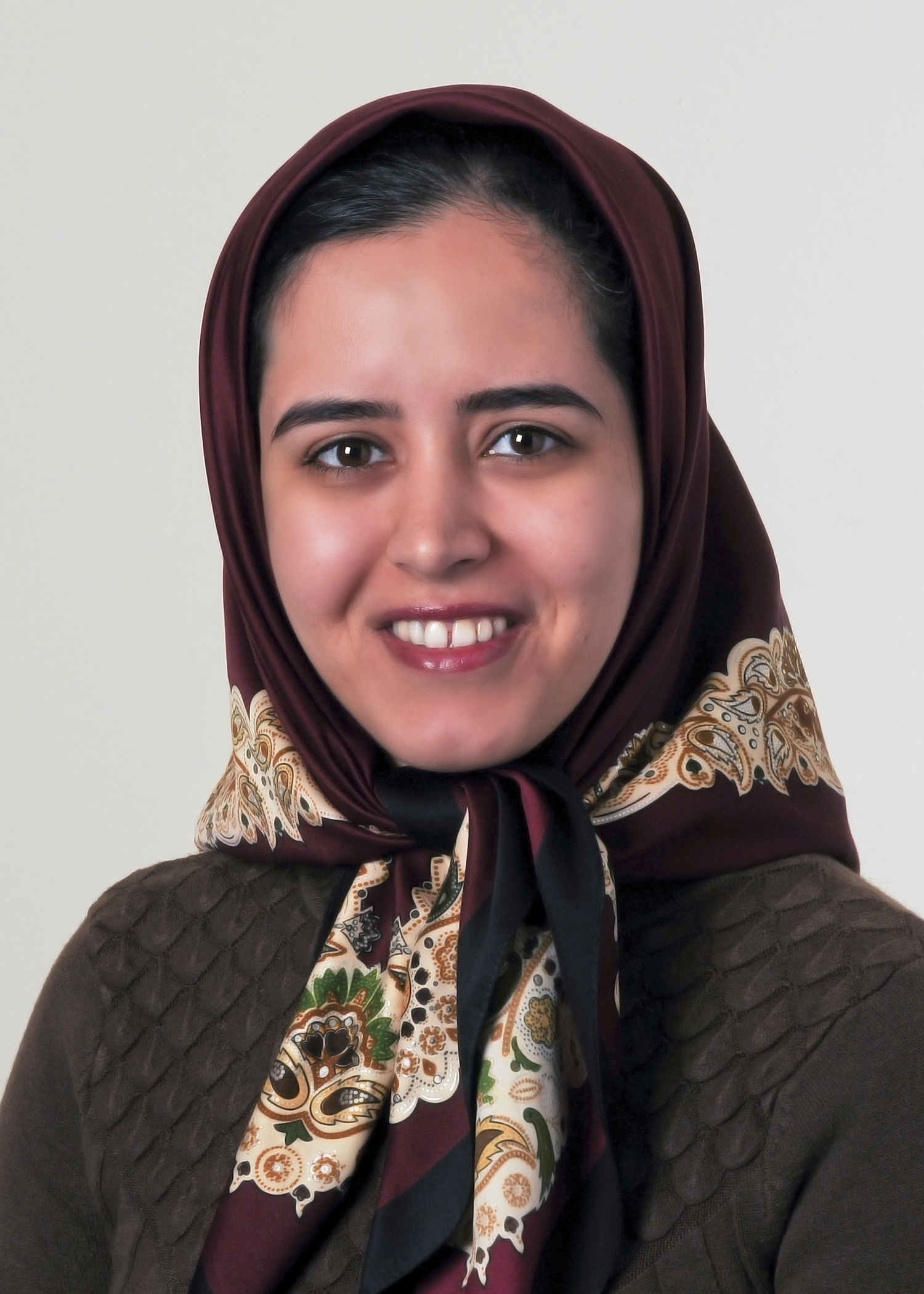}}]{Leila Karimi}
    received an undergraduate degree and the MS degree in information technology engineering from the Sharif University of Technology, Tehran, Iran. She is a Ph.D. candidate at the School of Computing and Information (SCI), at the University of Pittsburgh. Her research interests lie at the intersection of information security, data privacy, and machine learning. Currently, she is working on applying machine learning techniques to solve challenging problems in the security domain.  
\end{IEEEbiography}
\begin{IEEEbiographynophoto}{Maryam Aldairi}
    received an undergraduate degree management information systems From King Faisal University, Alhasa, KSA., and the MS degree in information science from the University of Pittsburgh. She is a Ph.D. student at the School of Computing and Information (SCI), at the University of Pittsburgh. Her research interests lie at the intersection of information security, adversarial learning, and machine learning. Currently, her focus is on applying machine learning techniques to solve challenging problems in the security domain.
\end{IEEEbiographynophoto}
\begin{IEEEbiography}
[{\includegraphics[width=1in,height=1.25in,clip]{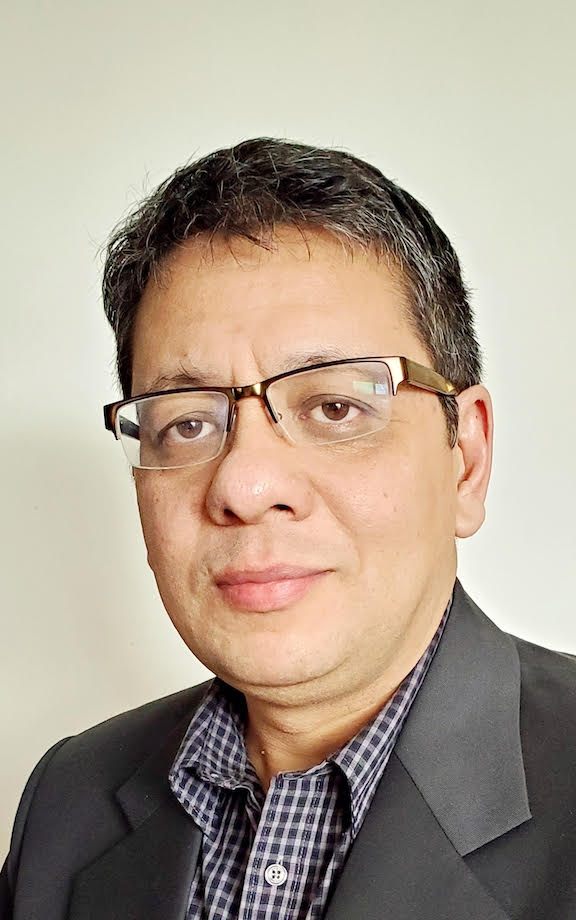}}]{James Joshi} received the MS degree in computer science and the Ph.D. degree in computer engineering from Purdue University. He is a professor of School of Computing and Information (SCI), at the University of Pittsburgh. His research interests include Access Control Models, Security and Privacy of Distributed Systems, Trust Management and Information Survivability. He is the director of LERSAIS at the University of Pittsburgh. He is an elected fellow of the Society of Information Reuse and Integration (SIRI) and is a senior member of the IEEE and the ACM. He currently serves as a Program Director of the Secure and Trustworthy Cyberspace program at the National Science Foundation.
\end{IEEEbiography}
\begin{IEEEbiography}
[{\includegraphics[width=1in,height=1.25in,clip]{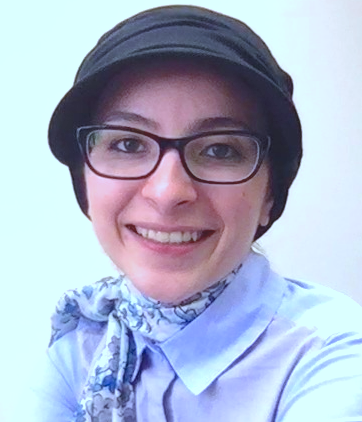}}]{Mai Abdelhakim} is an assistant professor in the department  of  electrical  and  computer  engineering at  the  University  of  Pittsburgh’s  Swanson school of engineering. She received her Ph.D. degree in Electrical Engineering from Michigan State University, and Bachelor’s and Master’s degrees in Electronics and Communications Engineering from Cairo University. Her research interests include cyber-physical systems, cybersecurity, machine learning, stochastic systems modeling, and information theory.
\end{IEEEbiography}




\end{document}